\documentclass[graybox]{svmult}
\usepackage{mathptmx}
\usepackage{graphicx}
\usepackage[bottom]{footmisc}
\usepackage{helvet}
\usepackage{courier}
\usepackage{type1cm}
\usepackage{multicol}

\usepackage{mathptmx}       % selects Times Roman as basic font
\usepackage{helvet}         % selects Helvetica as sans-serif font
\usepackage{courier}        % selects Courier as typewriter font
\usepackage{type1cm}        % activate if the above 3 fonts are
\usepackage{makeidx}         % allows index generation
\usepackage{graphicx}        % standard LaTeX graphics tool
\usepackage{multicol}        % used for the two-column index
\usepackage[bottom]{footmisc}% places footnotes at page bottom

\makeindex             % used for the subject index
\begin{document}

\title*{Magnetism in Dense Quark Matter}
\author{Efrain J. Ferrer and Vivian de la Incera}
\institute{Efrain J. Ferrer \at Department of Physics, University of Texas at El Paso, El Paso, Texas
79968, USA, \email{ejferrer@utep.edu}
\and Vivian de la Incera \at Department of Physics, University of Texas at El Paso, El Paso, Texas
79968, USA \email{vincera@utep.edu}}
\maketitle

\abstract{We review the mechanisms via which an external magnetic field can affect the ground state of cold and dense quark matter. In the absence of a magnetic field, at asymptotically high densities, cold quark matter is in the Color-Flavor-Locked (CFL) phase of color superconductivity characterized by three scales:  the superconducting gap, the gluon Meissner mass, and the baryonic chemical potential. When an applied magnetic field becomes comparable with each of these scales, new phases and/or condensates may emerge.  They include the magnetic CFL (MCFL) phase that becomes relevant for fields of the order of the gap scale; the paramagnetic CFL, important when the field is of the order of the Meissner mass, and a spin-one condensate associated to the magnetic moment of the Cooper pairs, significant at fields of the order of the chemical potential.  We discuss the equation of state (EoS) of MCFL matter for a large range of field values and consider possible applications of the magnetic effects on dense quark matter to the astrophysics of compact stars.}

\section{Introduction}
\label{Intro}

Contrary to what our na\"{\i}ve intuition might indicate, a magnetic field does not need to be of the order of the baryon chemical potential to produce a noticeable effect in a color superconductor. As discuss in \cite{SpinoneCFL,phases}, a color superconductor can be characterized by various scales and different physics can occur at field strengths comparable to each of them. Specifically,  for the so-called Color-Flavor-Locked (CFL) phase, the superconducting gap, the Meissner mass of the charged gluons, and the baryon chemical potential define three scales that determine the values of the magnetic field needed to produce different effects. Thus, the presence of sufficiently strong fields can modify the properties of the dense-matter phase which in turn might lead to observable signatures.

In this paper we review the status of our current knowledge about the magnetic field effects on color superconductivity (CS) at asymptotically high densities and discuss possible consequences of these effects for the physics of compact stars.

\section{Magnetic Fields in Compact Stars }
\label{sec:1}
The density of matter in the core of compact stars is expected to exceed that of nuclear matter $\rho_{nuc}=2.8\times10^{14}$ $g/cm^3$ \cite{Shapiro}. At such densities, individual nucleons overlap substantially. Under such conditions matter might consist of weakly interacting quarks rather than of hadrons \cite{Collins}. Due to the asymptotic freedom mechanism  \cite{AF-1,AF,AF-2} one might think that at  high baryon density QCD is amenable to perturbative techniques \cite{Perturbative-1,Perturbative-2,Perturbative-3,Perturbative-4,Perturbative-5,Perturbative-6,Perturbative-7,Perturbative-8}. However, the ground state of the superdense quark system, a Fermi liquid of weakly interacting quarks, is unstable with respect to the formation of diquark condensates \cite{CS-1-3,CS-1-2, CS-1}, a non-perturbative phenomenon essentially equivalent to the Cooper instability of BCS superconductivity. Given that in QCD one gluon exchange between two quarks is attractive in the color-antitriplet channel, at sufficiently high density and sufficiently small temperature quarks should condense into Cooper pairs, which are color antitriplets. These color condensates break the SU(3) color gauge symmetry of the ground state producing a color superconductor. 

In the late 90's the interest in CS was regained after the finding,  based on different effective theories for low energy QCD \cite{CS-2,CS-2-2}, that a color-breaking diquark condensate of much larger magnitude than originally thought may exist already at relatively moderate densities (of the order of a few times the nuclear matter density). At densities much higher than the masses of the u, d, and s quarks, one can assume the three quarks as massless. In this asymptotic region the most favored state is the CFL phase \cite{CS-2}, characterized by a spin-zero diquark condensate antisymmetric in both color and flavor. Since the combination of high densities and relatively low temperatures can naturally exist in the dense cores of compact stars, it is expected that CS could be realized in that astrophysical environment.

Compact stars, on the other hand, are very magnetized objects. From the measured periods and spin down of soft-gamma repeaters (SGR) and anomalous X-ray pulsars (AXP), as well as the observed X-ray luminosities of AXP, it has been found that a certain class of neutron stars named magnetars can have surface magnetic fields as large as $10^{14}-10^{16}$ G \cite{Magnetars-1, Magnetars-2, Magnetars-3, Magnetars-4, Magnetars-5}. In addition, since the stellar medium has a very high electric conductivity, the magnetic flux should be conserved. Hence, it is natural to expect an increase of the magnetic field strength with increasing matter density, and consequently a much stronger magnetic field in the stars' core. Nevertheless, the interior magnetic fields of neutron stars are not directly accessible to observation, so one can only estimate their values with heuristic methods. Estimates based on macroscopic and microscopic analysis, for nuclear \cite{virial-2, virial-1}, and quark matter considering both gravitationally bound and self-bound stars \cite{H-Estimate}, have led to maximum fields within the range $10^{18}-10^{20}$ G, depending whether the inner medium is formed by neutrons \cite{virial-2, virial-1}, or quarks \cite{H-Estimate}.

For instance, from energy-conservation arguments we can estimate the maximum field strength for a quark star. One should expect that the magnetic energy density does not exceed the energy density of the self-bound quark matter, which is given as the energy density at zero pressure that can have a maximum value equal to that of the iron nucleus (roughly 939 MeV). Based on this reasoning, the maximum field allowed can be estimated as
\begin{equation}\label{Bind-Energy}
B_{max}\simeq \frac{\varepsilon_{bind}^2}{e\hbar c}\leq \frac{(939 MeV)^2}{e\hbar c}\sim1.5\times10^{20} G,
\end{equation}
From this result, one notice that the inner field can reach values two orders of magnitude larger than the estimates done for gravitationally bound stars assuming that they contain nuclear matter \cite{virial-2, virial-1}.

As we will see in Sec. 8, the magnetic field decreases the inner pressure along the field direction of the magnetized system. If the magnetic field that makes such a pressure component equal to zero is taken as the maximum value of the inner field allowable for a stable gravitational bound star, then a star with a quark matter core can have a maximum field $\sim 10^{19}-10^{20}$ G \cite{EoS-H}, while one with nuclear matter can only have fields  $\sim 10^{18}$ G \cite{virial-2, virial-1}.

Therefore, the investigation of the properties of very dense matter in the presence of strong magnetic fields is of interest not just from a fundamental point of view, but it could be also closely connected to the physics of strongly magnetized neutron stars.

\section{Magnetism in Spin-Zero Color Superconductivity}
\label{sec:2}

An important point to keep in mind in our analysis of the field effects in CS is that in spin-zero color superconductivity the electromagnetism is not the conventional one. In the color superconducting medium the conventional electromagnetic field is not an actual eigenfield, since it is mixed with one of the gluon fields, much like the mixing ocurring between the hyper-field and the W-boson in the electroweak model in the presence of the Higgs condensate. Thus, even though the original electromagnetic $U(1)_{em}$ symmetry is broken by the formation of the charged quark Cooper pairs in the CFL phase \cite{CFL}, a residual $\widetilde{U}(1)$ gauge symmetry still remains. The massless gauge field associated with this symmetry is given by the linear combination of the conventional photon field and the $8^{th}$ gluon field \cite{CFL, ABR, ABR-1},
\begin{equation}\label{rotated field}
 {\tilde A}_\mu=\cos\theta A_\mu-\sin\theta G^8_\mu.
\end{equation}
The corresponding orthogonal linear combination
\begin{equation}
\tilde G_{\mu}^8=\sin\theta A_{\mu}+\cos\theta G_{\mu}^8.
\end{equation}
is massive. The field $\tilde A_{\mu}$ plays the role of an in-medium or rotated electromagnetic field. A magnetic field associated with $\tilde A_{\mu}$ can penetrate the CS without being subject to the Meissner effect, since the color condensate is neutral with respect to the rotated charge. However, the rotated electromagnetic field in the CFL superconductor is mostly formed by the original photon with only a small admixture of the $8^{th}$ gluon since the mixing angle, $\cos\theta=g/\sqrt{e^2/3+g^2}$, is sufficiently small.

The generator of the unbroken $\widetilde U(1)$ symmetry, which corresponds to the long-range rotated photon in the CFL phase, is a matrix in $flavor_{(3\times3)}\otimes color_{(3\times3)}$ space given by ${\tilde Q}_{CFL}=Q\otimes I+I\otimes T_8/\sqrt{3}$, where $Q$ is the conventional electromagnetic charge operator of quarks and $T_8$ is the $8^{th}$ Gell-Mann matrix. Using the matrix representations, $Q=diag(-1/3,-1/3,2/3)$ for $(s,d,u)$ flavors, and $T_8=diag(-1/\sqrt{3},-1/\sqrt{3},2/\sqrt{3})$ for $(b,g,r)$ colors, the ${\tilde Q}$ charges (in units of ${\tilde e}=e\cos\theta$) of different quarks are
\begin{equation}\label{table-CFL}
\begin{tabular}{ccccccccc}
\hline
\textrm{$s_b$}&
\textrm{$s_g$}&
\textrm{$s_r$}&
\textrm{$d_b$}&
\textrm{$d_g$}&
\textrm{$d_r$}&
\textrm{$u_b$}&
\textrm{$u_g$}&
\textrm{$u_r$}\\
0 & 0 & $-1$ & 0 & 0 & $-1$ & $+1$ & $+1$ & 0\\
\hline
\end{tabular}
\end{equation}

For the $2SC$ color superconductor the ground state is formed by spin-zero diquarks, which are also neutral with respect to the rotated electromagnetic field associated with the remnant $\widetilde{U}(1)$ symmetry. In this system (see Refs. \cite{CS-2, CS-1-3, Casalbuonu, CS-2-2},
the generator of the remnant $\widetilde U(1)$ symmetry is a matrix in $flavor_{(2\times2)}\otimes color_{(3\times3)}$ space given by ${\tilde Q}_{2SC}=Q\bigotimes I-I\bigotimes T_8/\sqrt{3}$, with the usual matrix of electromagnetic charges of quarks in flavor space $Q=diag(2/3,-1/3)$, and $T_8$ is the eighth generator
of the $SU(3)_c$ gauge group in the adjoint representation. The rotated charges of the quarks in units of $\widetilde{e} =e\cos \theta$, are given in this phase by
\begin{equation}\label{table-2SC}
\begin{tabular}{cccccc}
\hline
\textrm{$d_b$}&
\textrm{$d_g$}&
\textrm{$d_r$}&
\textrm{$u_b$}&
\textrm{$u_g$}&
\textrm{$u_r$}\\
$ -\frac{1}{2}$ &$ -\frac{1}{2} $& $0$ & $\frac{1}{2}$ & $\frac{1}{2}$ &1\\
\hline
\end{tabular}
\end{equation}
and the massless rotated electromagnetic field and orthogonal massive field are defined in the same way as in the CFL case.

From now on, we will use "magnetic field" in short, when we refer to the "rotated magnetic field," since inside the superconductor only the rotated magnetic field is the physical long range field.

\section{The Magnetic CFL Phase}
\label{sec:3}

The fact that the rotated magnetic field can penetrate the spin-zero color superconductor brings the possibility to look for possible field-interaction effects on the CS phase. An important consequence of this interaction was first studied in \cite{MCFL}. It is based in the following observation. Although the Cooper pairs have zero rotated charge, they can be formed either by neutral quarks or by quarks of opposite charges. If the magnetic field is strong enough so that the magnetic length $l_0=1/\sqrt{2eB}$ becomes smaller than the pairs' coherence length, then the magnetic field can interact with the pair constituents and significantly modify the pair structure of the condensate. As shown in \cite{MCFL,MCFL-1,MCFL-2,MCFL-3}, the presence of a magnetic field changes the CFL phase characterized by one single gap, producing a splitting of the CFL gap into a gap that gets contributions from both pairs of oppositely charged, as well as neutral, quarks, denoted by $\Delta_{B}$, and one that only gets contributions from pairs of neutral quarks, denoted by $\Delta$. The new phase that forms in the presence of the magnetic field also has color-flavor-locking, but with a smaller symmetry group $SU(2)_{C+L+R}$, a change that is reflected in the splitting of the $\Delta$ and $\Delta_B$ gaps.
The less symmetric realization of the CFL pairing that occurs in the presence of a magnetic field, is known as the magnetic-CFL (MCFL) phase \cite{MCFL,MCFL-1,MCFL-2,MCFL-3}. The MCFL phase has similarities, but also important differences with the CFL phase \cite{MCFL,MCFL-1,MCFL-2,MCFL-3,phases,MCFLoscillation,MCFLoscillation-1}.

In the strong-field limit, the gap formed by pairs of neutral and charged quarks satisfies the gap equation \cite{MCFL,MCFL-1,MCFL-2,MCFL-3}
\begin{equation} \label{maingeq}
1  \approx  \frac{g^2}{3 \Lambda^2}
\int_{\Lambda} \frac{d^3 q}{(2 \pi)^3} \frac{
1}{\sqrt{(q-\mu)^2 + 2 (\Delta_B)^2 }}
 +  \frac{g^2 \widetilde{e}\widetilde{B}}{6 \Lambda^2} \int_{-\Lambda}^{\Lambda}
\frac{d q}{(2 \pi)^2} \frac{ 1}{\sqrt{(q-\mu)^2 +
(\Delta_B)^2 }}   \ , \end{equation}

The solution of (\ref{maingeq}) is given by
\begin{equation} \label{deltaB}
\label{gapBA} \Delta_B \sim 2 \sqrt{\delta \mu} \, \exp{\Big( -
\frac{3 \Lambda^2 \pi^2} {g^2 \left(\mu^2 + \frac{\widetilde{e}
\widetilde{B}}{2} \right)} \Big) } \ ,
\end{equation}
 which can be compared with the
CFL gap
\begin{equation}
\label{CFL-Agap} \Delta_{\rm CFL} \sim 2
\sqrt{\delta \mu} \,\exp{  \Big(-\frac{ 3 \Lambda^2 \pi^2}{ 2 g^2
\mu^2}\Big)} \ ,
\end{equation}
Here we used $\delta \equiv \Lambda - \mu$, with  $\Lambda$  the ultraviolet cutoff of the NJL model that should be much
larger than any of the typical energy scales of the system, and $\mu$ the baryon chemical potential.

The gap $\Delta$, formed only by pairs of neutral quarks, should be found as the solution of the gap equation
\begin{equation} \label{antisymme-gapeqapp}
 1  \approx  \frac {g^2}{4 \Lambda^2}  \int_{\Lambda} \frac{d^3q}{(2 \pi)^3}
 \Big( \frac {17}{9} \frac{1}{ \sqrt{ (q - \mu)^2 - \Delta^2 }}
 +   \frac{7}{9} \frac{1}{ \sqrt{(q-\mu)^2 + 2 (\Delta_B)^2 }  }
 \Big) \ ,
 \end{equation}
where it is apparent the interconnection with the gap $\Delta_B$. This is how through $\Delta_B$ the magnetic field can affect $\Delta$ although it is formed only by neutral quarks as we already pointed out.

The solution of (\ref{antisymme-gapeqapp}) is given by
\begin{equation}
 \Delta \sim \frac{1}{2^{(7/34)}} \exp{\Big(-\frac{36}{17 x}  + \frac{21}{17}\frac{1}{x (1 + y)} + \frac{3}{2x}
\Big)\Delta_{\rm CFL} } \ ,
\end{equation}
where $x \equiv g^2 \mu^2/\Lambda^2 \pi^2$, and $y \equiv
\widetilde{e}\widetilde{B}/\mu^2$

The exponent in
(\ref{gapBA}) has the typical BCS form, but with different density
of states for neutral and charged quarks, i.e. $exp ~[~1/(N_{\mu}
+N_{\widetilde{B}})~ \widetilde{G}~]$, where $N_\mu=\mu^2/\pi^2$ is
the density of states at the Fermi surface of the neutral quarks
with single chirality,
$N_{\widetilde{B}}=\widetilde{e}\widetilde{B}/2\pi^2$ is the density
of states of the charged quarks lying at the zero Landau level at the Fermi surface, and $\widetilde{G}=-g^2/3\Lambda^{2}$ is the
characteristic effective coupling constant of the $\overline{3}$
channel \cite{Son}. The effect of the strong magnetic field 
$\widetilde{e}\widetilde{B}/2 \geq \mu^2$ is to increase the total
density of states, thus producing a gap enhancement.

Although the situation here shares
some similarities with the magnetic catalysis of chiral symmetry
breaking \cite{ MC-7, MC-8, MC-AMM, MC-AMM-2, MC-1, MC-2, MC-3, MC-4, MC-5, MC-6}; the way the field influences the pairing
mechanism in the two cases is quite different. The particles
participating in the chiral condensate are near the surface of the
Dirac sea. The effect of a magnetic field there is to effectively
reduce the space dimension where the particles are embedded at the lowest Landau level (LLL),
which as a consequence strengthens their effective coupling, and so catalyzing the
chiral condensate. Color superconductivity, on the other hand,
involves quarks near the Fermi surface, with a pairing dynamics that
is already $(1+1)$-dimensional. Therefore, the ${\widetilde B}$
field does not yield further dimensional reduction of the pairing
dynamics near the Fermi surface and hence the LLL
does not have a special significance here. Nevertheless, the field
increases the density of states of the ${\widetilde Q}$-charged
quarks, and it is through this effect, as shown in Eq.
(\ref{gapBA}), that the pairing of the charged particles is
reinforced by the penetrating magnetic field.

Note that our analytic solutions are only valid at strong magnetic
fields.  For fields of this order and larger, the $\Delta_B$ gap is
larger than $\Delta_{\rm CFL}$ at the same density values. How
fast or slow the gaps do it depends very much on the values of the NJL
couplings. For example, for $x \sim 0.3$,
one finds $\Delta \sim 0.2\, \Delta_B$ for $y= 3/2$, while for
$x \sim 1$ then $\Delta \sim 0.5 \,\Delta_B$.

In a recent study \cite{SpinoneCFL}, it was discovered that the MCFL phase actually contains one more condensate, which we will call $\Delta_M$. This new condensate is associated with the magnetic moment of the Cooper pairs. Physically this is easy to understand. The presence of a uniform magnetic field explicitly breaks the spatial rotational symmetry $O(3)$ to the subgroup $O(2)$ of rotations about the axis along the field. As shown in \cite{SpinoneCFL}, this symmetry reduction has non-trivial consequences for the ground state structure of the MCFL superconductor. When one performs the Fierz transformations in the quark system with both Lorentz and rotational $O(3)$ symmetries explicitly broken, various new pairing channels appear allowing in principle the formation of new condensates. Of particular interest is an attractive channel that leads to a spin-one condensate of Dirac structure $\Delta_M\sim C\gamma_5\gamma^1\gamma^2$. Such a gap does not break any symmetry that has not already been broken by the other condensates of the MCFL ground state, so it in principle is not forbidden. The new condensate corresponds to the zero spin projection of the average magnetic moment of the Cooper pairs in the medium.

From a physical point of view, it is natural to expect the formation of this extra condensate in the magnetized system because the diquarks formed by oppositely charged quarks with opposite spins will have a net magnetic moment that may point parallel or antiparallel to the magnetic field.  Diquarks formed by quarks lying on any non-zero Landau level can have magnetic moments pointing in both directions, because each quark in the pair may have both spins. Hence the contribution of these diquarks to the net magnetic moment should tend to cancel out. On the other hand,  diquarks from quarks in the LLL can only have one orientation of their magnetic moment with respect to the field, because the quarks in the LLL have only one possible spin projection. This implies that the main contribution to the new condensate should come from the quarks at the LLL, an expectation that is consistent with the numerical results found in \cite{SpinoneCFL}, where the new gap was obtained to be negligibly small at weak magnetic fields, where the zero Landau level occupation is not significant. On the other hand, at strong magnetic fields, the condensate became comparable in magnitude to the original condensates, $\Delta$ and $\Delta_B$, of the MCFL ground state \cite{MCFL}, because the majority of the quarks occupy the LLL in that case.

Although this new condensate is zero at zero magnetic field, we cannot ignore it even at very small magnetic fields because a self-consistent solution of the gap equations with $\Delta\neq 0$, and $\Delta_B\neq 0$, but  $\Delta_M=0$ is not possible. This is easy to understand since, as long as the magnetic field is not zero, there is always some occupation of the LLL. Thus, once a magnetic field is present,  $\Delta_M$ has to be considered simultaneously with the spin-zero MCFL gaps.

The $\Delta_M$ condensate of the  MCFL scenario described above shares a few similarities with the dynamical generation of an anomalous magnetic moment recently found in massless QED \cite{MC-AMM, MC-AMM-2}. Akin to the Cooper pairs of oppositely charged quarks in the MCFL phase, the fermion and antifermion that pair in massless QED also have opposite charges and spins and hence carries a net magnetic moment. A dynamical magnetic moment term in the QED Lagrangian does not break any symmetry that has not already been broken by the chiral condensate. Therefore, once the chiral condensate is formed due to the magnetic catalysis of chiral symmetry breaking \cite{ MC-7, MC-8, MC-1, MC-2, MC-3, MC-4, MC-6, MC-5}, the simultaneous formation of a dynamical mass and a dynamical magnetic moment is unavoidable \cite{MC-AMM, MC-AMM-2}. The realization of the anomalous magnetic moment condensate in magnetized massless QED produces a non-perturbative Zeeman effect \cite{MC-AMM, MC-AMM-2}.

At moderate magnetic fields the energy gaps $\Delta$ and $\Delta_B$ exhibit oscillations when ${\tilde e}{\tilde B}/\mu^2$ is varied \cite{MCFLoscillation,MCFLoscillation-1}, owed to the de Haas-van Alphen effect \cite{HvA, HvA-1} typical of charged fermion systems under magnetic fields (see for instance \cite{Klimenko, Klimenko-1,MCFLoscillation,MCFLoscillation-1}), while for $\Delta_M$ the oscillations are almost absent \cite{SpinoneCFL}. These features indicate, as already pointed out, that the main contribution to $\Delta_M$ should come from pairs whose charged quarks are at the LLL.

\begin{figure}
\begin{center}
\includegraphics[width=0.7\textwidth]{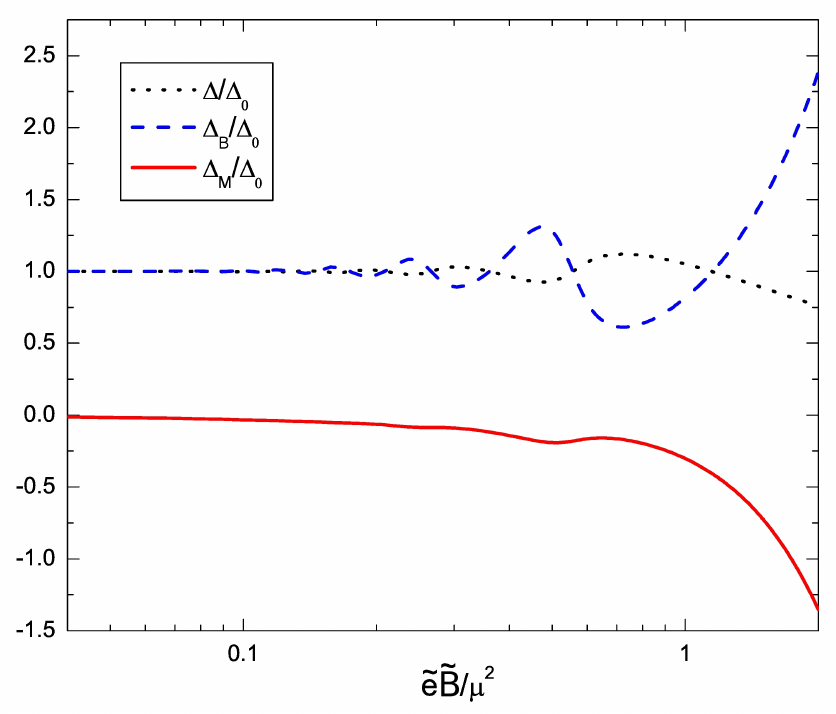}
\caption{The three gaps of the MCFL phase as a function of ${\tilde e}{\tilde B}/\mu^2$ for $\mu=500MeV$. They are scaled with respect to the CFL gap $\Delta_{CFL}=25MeV$.}
\label{fig1}
\end{center}
\end{figure}

The previous discussion can be visualized in the plot of the gaps as functions of a dimensionless parameter ${\tilde e}{\tilde B}/\mu^2$ given in Fig.\ref{fig1}. Note that for small magnetic field, $\Delta$ and $\Delta_B$ are close to each other and approach the CFL gap $\Delta_{CFL}=25MeV$. As the magnetic field increases, $\Delta$ and $\Delta_B$  display oscillatory behaviors with respect to ${\tilde e}{\tilde B}/\mu^2$ as long as ${\tilde e}{\tilde B}<\mu^2$.
As originally explained by Landau \cite{Lev}, these oscillations reveal the quantum nature of the interaction of the charged particles with the magnetic field (the well-known Landau quantization phenomenon), and are produced by the change in the density of states when passing from one Landau level to another. The oscillations cease when the first Landau level exceeds the Fermi surface. For ultra-strong fields, when only the LLL contributes to the gap equation, $\Delta_B$ is much larger than $\Delta$, as it was found by analytical calculation in \cite{MCFL}.

The only contribution to $\Delta_M$ from higher LLs can come when the number of particles is odd, so there are energy states occupied by a single particle, but that is a very small part. The cancelation does not occur, however, between the pairs of quarks in the LLL because they can only be formed by positive quarks with spin up and negative quarks with spin down. At low fields, the number of quarks in the LLL is scarce, while for fields of order ${\tilde e}{\tilde B}\geq \mu^2$, all the particles are constrained to the LLL, hence the variation of $\Delta_M$ from lower values at weak field, to higher values at sufficiently strong fields.

It is apparent from the graphical representation of $\Delta_M$ in Fig.{\ref{fig1}}, that its value remains relatively small up to magnetic-field values of the order of $\mu^2$. In the field region between $10^{18}-10^{19}G$, the magnitude of $\Delta_M$ grows from a few tenths of Mev to tens of Mev. It becomes comparable to the MCFL gap $\Delta_{B}$ when the field is strong enough to put all the quarks in the LLL, shown in the final segment of the plots in the figure.

Another important consequence of the gap $\Delta_M$ is the increment in the magnitude of $\Delta_{B}$ for any given value of the magnetic field in the strong field region, as compared to its own value found at the same field but ignoring the existence of $\Delta_M$ \cite{MCFLoscillation,MCFLoscillation-1}. This effect, combined with the increase of $\Delta_M$ at strong fields, will make the MCFL phase more stable than the regular CFL, a fact that could favor the realization of an MCFL core in magnetars.

Let us now discuss in more detail the difference between the CFL and MCFL phases from the point of view of symmetry. In the absence of a magnetic field,
three-flavor massless quark matter at high baryonic density is in
the energetically favored CFL phase. There, the diquark condensates lock the color and
flavor transformations, breaking both symmetries. Thus, the symmetry
breaking pattern in the CFL phase is
\begin{equation}\label{CFL}
 SU(3)_C \times SU(3)_L \times SU(3)_R \times U(1)_B
\rightarrow SU(3)_{C+L+R} \ .
\end{equation}
In this case, there are only nine Goldstone bosons that survive to the
Anderson-Higgs mechanism. One is a singlet, scalar mode,
associated to the breaking of the baryonic symmetry, and the
remaining octet is associated to the axial $SU(3)_A$ group, just
like the octet of mesons in vacuum. At sufficiently high density, the anomaly is
suppressed, and then one can as well consider the spontaneous breaking of an approximated
$U(1)_A$ symmetry, and the additional pseudo
Goldstone boson. We will ignore this effect, though.

Once electromagnetic effects are considered, the flavor symmetries
of QCD are reduced, as only the $d$ and $s$ quarks have equal
electromagnetic charges, $q= -e/3$, while the $u$ quark has
electromagnetic charge, $q=2e/3$. However, because the
electromagnetic structure constant $\alpha_{\rm e.m.}$ is relatively
small, this effect is considered to be really tiny, a small
perturbation, and one can consider good approximated flavor symmetries.
Nevertheless, in the presence of a strong magnetic field one cannot consider the effects of electromagnetism as a small perturbation.
Flavor symmetries are explicitly reduced from $SU(3)_{L,R}$  to  ${SU(2)}_{L,R}$.
For sufficiently strong magnetic fields, in the MCFL phase
the symmetry breaking pattern is then
\begin{equation}\label{MCFL}
SU(3)_C \times SU(2)_L \times SU(2)_R \times U(1)^{(1)}_A\times U(1)_B \times U(1)_{\rm e.m.}
\rightarrow SU(2)_{C+L+R} .
\end{equation}
Here the symmetry group $U(1)^{(1)}_A$ is related to a current,
which is an anomaly free linear combination of $u,d$ and $s$ axial
currents, and such that  $U(1)^{(1)}_A \subset SU(3)_A$. The locked $SU(2)$ group
corresponds to the maximal unbroken symmetry, such that it maximizes the condensation
energy. The counting of broken generators, after taking into account the
Anderson-Higgs mechanism,  leads to only five Goldstone
bosons. As in the CFL case, one is associated with the breaking of
the baryon symmetry; three Goldstone bosons are associated with the
breaking of $SU(2)_A$, and another one with the breaking
of  $U(1)^{(1)}_A$. As before, if the effects of the anomaly could
be neglected, there would be another pseudo Goldstone
boson associated with the $U(1)_A$ symmetry.
Thus,  apart from modifying the value
of the gaps, an applied strong  magnetic field also affects the number of
Goldstone bosons, reducing them from nine (neutral and charged) to five (neutral).

Once a magnetic field is
present, the original symmetry group is reduced, and the low energy theory correspond to the breaking pattern (\ref{MCFL}), hence be
described by five Goldstone bosons. In practice however, at weak magnetic fields, it is reasonable to treat
the symmetry of the CFL phase as a good approximated symmetry, which means that at
weak fields the low-energy excitations are essentially governed by
nine approximately massless scalars (those of the breaking pattern
(\ref{CFL})) instead of five.

A question of order here is: what do we exactly understand  as a
weak magnetic field? In other words, what is the threshold-field
strength that effectively separates the CFL low energy behavior from
the MCFL one? A fundamental clue in this direction was found in \cite{phases} by determining the term in the low-energy CFL Lagrangian that
generates a field-induced mass for the charged Goldstone fields, so
disconnecting them from the low-energy dynamics at some field
strength and thereby effectively reducing the number of Goldstone
bosons from the nine of the CFL phase, to the five neutral ones of the MCFL.

The threshold field $\widetilde{B}_{MCFL}$ for the effective $CFL \rightarrow MCFL$ symmetry crossover was found to be
 \cite{phases}
\begin{equation}\label{Threshold-value}
\widetilde{e}\widetilde{B}_{MCFL}=\frac{4}{v_{\bot}^2}\Delta_{CFL}^2\simeq
12\Delta_{CFL}^2.
\end{equation}
where the weak-field approximation $v_{\bot}\simeq
1/\sqrt{3}$ was considered \cite{Stephanov}. The threshold field does not depend on the decay
constant $f_{\pi}$, therefore it depends on $\mu$ only through
$\Delta_{CFL}$. For $\Delta_{CFL} \sim 15 MeV$ one gets ${\tilde e}
{\tilde B}_{MCFL} \sim 10^{16}G$. At these field strengths, the
charged mesons decouple from the low-energy theory. When this
decoupling occurs, the five neutral Goldstone bosons (including the
one associated to the baryon symmetry breaking) that characterize
the MCFL phase will drive the low-energy physics of the system.
Therefore, coming from low to higher fields, the first magnetic
phase that will effectively show up in the magnetized system for $\widetilde{e}\widetilde{B}\sim \Delta_{CFL}^2$ will be
the MCFL \cite{MCFL,MCFL-1,MCFL-2,MCFL-3}.

Summarizing, in a color superconductor with three-flavor quarks at
very high densities an increasing magnetic field produces a phase
crossover from CFL to MCFL. During this phase transmutation no symmetry
breaking occurs, since in principle once a magnetic field is present
the symmetry is strictly speaking that of the MCFL, as discussed above.
However, in practice for $\widetilde{B}\sim \widetilde{B}_{MCFL}\sim
\Delta^{2}_{CFL}$ the main features of MCFL emerge through the
low-energy behavior of the system \cite{phases}. At the threshold field
$\widetilde{B}_{MCFL}$, only five neutral Goldstone bosons remain
out of the original nine characterizing the low-energy behavior of
the CFL phase, because the charged Goldstone bosons acquire field dependent masses and can decay in lighter modes.
For a meson to be stable in this system, its mass
should be less than twice the gap, otherwise it will decay into a
particle-antiparticle pair. That means that, as proved in Ref. \cite{phases}, once the applied field produces a mass for the charged Goldstones of the order of the CFL gap it is reached the threshold field
for the effective $CFL \rightarrow MCFL$ symmetry transmutation.

The existence of this phase transmutation is on the other hand manifested in the behavior of the gaps versus the magnetic field. At field strength smaller than the threshold field we find that $\Delta \approx \Delta_B \approx \Delta_{CFL}$, while for fields closer to $\widetilde{B}_{MCFL}$ the gaps exhibit oscillations with respect to ${\tilde e}{\tilde B}/\mu^2$ \cite{SpinoneCFL,MCFLoscillation,MCFLoscillation-1}, owed to the de Haas-van Alphen effect \cite{HvA, HvA-1}.

It is worth to call attention to the analogy between the CFL-MCFL
crossover and what could be called a "field-induced" Mott transition. Mott transitions
were originally considered in condensed matter in the context of metal-insulator transitions
in strongly-correlated systems \cite{Mott, Mott-1}. Later on, Mott transitions have been also discussed in
QCD to describe delocalization of bound states into their constituents at a temperature
defined as the Mott temperature \cite{Hufner}. By definition, the Mott temperature $T_M$ is the temperature
at which the mass of the bound state equals the mass of its constituents, so the
bound state becomes a resonance at $T > T_M$. In the present work, the role of the Mott
temperature is played by the threshold field $\widetilde{B}_{MCFL}$. Mott transitions typically lead to the
appearance of singularities at $T = T_M$ in a number of physically relevant observables. It is
an open question, worth to be investigated whether similar singularities are or not present
in the CFL-MCFL crossover at $\widetilde{B}_{MCFL}$.

\section{Magnetoelectric Effect in Cold-Dense Matter}
\label{Magnetoelectric}

It is well known that the phenomenon of CS shares many characteristics of condensed matter systems \cite{CS}. In this section, we discuss a new feature of CS that has its counterpart in magnetically ordered materials and has been known in the context of condensed matter for many years. It is the so called magnetoelectric (ME) effect, which establishes a relation between the electric and magnetic properties of certain materials. In general, it states that the electric polarization of such materials may depend on an applied magnetic field and/or that the magnetization may depend on an applied electric field. The first observations of magnetoelectricity took place when a moving dielectric was found to become polarized when placed in a magnetic field \cite{early-ME-1, early-ME-2}. In 1894, Pierre Curie \cite{Curie} was the first in pointing out the possibility of an intrinsic ME effect for certain (non-moving) crystals on the basis of symmetry considerations. But it took many decades to be understood and proposed by Landau and Lifshitz \cite{L-L} that the linear ME effect is only allowed in time-asymmetric systems. Recently the ME effect regained new interest in condensed matter thanks to new advancements in material science and with the development of the so-called multiferroic materials for which the ME effect is significant for practical applications \cite{Revival}.

As demonstrated in Refs. \cite{EM, Magnetoelectric-CS} the ME effect also occurs in a highly magnetized CS medium like the MCFL phase. In particular, in \cite{EM, Magnetoelectric-CS} it was shown how the electric susceptibility of this medium depends on an applied strong magnetic field.

Let us start by discussing the ME effect at weak fields. At weak fields this effect can be studied by taking into account the expansion of the system's free energy in powers of the electric $\bf \widetilde{E}$ and magnetic $\bf \widetilde{B}$ fields
\begin{eqnarray}\label{N-1}
F({\bf \widetilde{E}},{\bf \widetilde{B}})=F_0-\alpha_i\widetilde{E}_i-\beta_i\widetilde{B}_i-\gamma_{ij}\widetilde{E}_i\widetilde{B}_j-\eta_{ij}\widetilde{E}_i\widetilde{E}_j-\tau_{ij}
\widetilde{B}_i\widetilde{B}_j\nonumber
\\
-\kappa_{ijk}\widetilde{E}_i\widetilde{E}_j\widetilde{B}_k-\lambda_{ijk}\widetilde{E}_i\widetilde{B}_j\widetilde{B}_k-\sigma_{ijkl}\widetilde{E}_i\widetilde{E}_j\widetilde{B}_k\widetilde{B}_l-\ldots
\end{eqnarray}

In this weak-field expansion the coefficients $\alpha_i, \gamma_{ij}$ etc., which are the susceptibility tensors, can be found from the infinite set of one-loop polarization operator diagrams with external legs of the in-medium photon field $\widetilde{A}_{\mu}$ and internal lines of the full CFL quark propagator of the rotated charged quarks. Hence, these coefficients can only depend on the baryonic chemical potential, the temperature and the CFL gap. From (\ref{N-1}), the electric polarization can be found as
\begin{equation} \label{Polarization}
P_i=-\frac{\partial F}{\partial \widetilde{E}_i}=\alpha_i+\gamma_{ij}\widetilde{B}_j+2\eta_{ij}\widetilde{E}_j+2\kappa_{ijk}\widetilde{E}_j\widetilde{B}_k+\lambda_{ijk}\widetilde{B}_j\widetilde{B}_k+2\sigma_{ijkl}\widetilde{E}_j\widetilde{B}_k\widetilde{B}_l+\ldots
\end{equation}

If the tensor $\gamma$ is different from zero the system exhibits the linear ME effect. From the free energy (\ref{N-1}), we see that the linear ME effect can only exist if the time-reversal and parity symmetries are broken in the medium. In the CFL phase, the time-reversal symmetry is broken by the CFL gap \cite{SpinoneCFL}, but parity is preserved. Thus, the linear ME effect cannot be present in this medium. The behavior under a time-reversal transformation underscores an important difference between the CFL color superconductivity and the conventional, electric superconductivity. While the CFL color superconductor is not invariant under time-reversal symmetry, the conventional superconductor is, since in the conventional superconductor the Cooper pairs are usually formed by time-reversed one-particle states \cite{Anderson}. In the conventional superconductor the violation of the T-invariance occurs only via some external perturbation which can lead in turn to pair breaking and to the so-called gapless superconductivity \cite{Abrikosov}.

Higher-order ME terms are parameterized by the tensors $\kappa$, and $\lambda$. As it happens with $\gamma$, the coefficient $\lambda \neq 0$ is forbidden because it requires parity violation. On the other hand, although a $\kappa \neq 0$ term only requires time-reversal violation, to form a third-rank tensor independent of the momentum and parity invariant, the medium would need to have an extra spatial vector structure. However, the only tensor structures available to form such a third-rank tensor in the CFL phase are the metric tensor $g_{\mu \nu}$  and the medium fourth velocity $u_\mu$, which in the rest frame is a temporal vector $u_\mu=(1,0,0,0)$, so the coefficient $\kappa$ should be zero too. Hence, we do not expect any ME effect associated with the lower terms in the weak-field expansion of the free energy (\ref{Polarization}).

At strong magnetic fields, the situation is quite different. In this case the expansion of the free energy can only be done in powers of a weak electric field, and the coefficients of each term can be found from the corresponding one-loop polarization operators, which now depend on the strong magnetic field in the MCFL phase. The free energy expansion in this case takes the form
\begin{equation}\label{free-energy-strong}
F'({\bf \widetilde{E}},{\bf \widetilde{B}})=F'_0(\widetilde{B})-\alpha'_i\widetilde{E}_i-\eta'_{ij}\widetilde{E}_i\widetilde{E}_j-\ldots
\end{equation}
The tensors $\alpha'$ and $\eta'$ can depend now on the baryonic chemical potential, temperature, magnetic field and gaps of the MCFL phase \cite{SpinoneCFL}. They can be found respectively by calculating the tadpole and the second rank polarization operator tensor of the MCFL phase in the strong field limit. An $\alpha'\neq 0$ would indicate that the MCFL medium behaves as a ferroelectric material \cite{ferroelectricity-1, ferroelectricity-2}, but this is not the case because this phase is parity symmetric \cite{SpinoneCFL}, hence $\alpha' = 0$. The tensor $\eta'$, nevertheless, is not forbidden by any symmetry argument. If it is different from zero, $\eta'$ would characterize the lowest order of the system dielectric response. More important, if $\eta'$ results to be dependent on the magnetic field, this would imply that the electric polarization $P=\eta' E$ depends on the magnetic field through $\eta'$, hence the MCFL phase would exhibit the ME effect.

To find the electric susceptibility $\eta'$ in the strong-magnetic-field limit of the MCFL phase we start from \cite{SpinoneCFL}
\begin{equation}\label{Strong-B-Suscept}
   F'({\bf \widetilde{E}},{\bf \widetilde{B}})-F'_0({\bf\widetilde{B}})\sim\frac{1}{V}\int \widetilde{A}_0(x_3)\Pi_{00}(x_3-x_3')\widetilde{A}_0(x_3')dx_3dx_3'=-\eta' \widetilde{E}^2,
\end{equation}
our task is then reduced to the calculation of the zero-zero component of the one-loop polarization operator at strong magnetic field in the infrared limit, $\Pi_{00}(p_0=0, p\rightarrow 0)$.

Now, the photon polarization operator should be gauge invariant. That is, in the strong-field approximation, it should satisfy the transversality condition in the reduced $(1+1)$-D space (${p_\mu}^\parallel\Pi_{\mu\nu}^\parallel(p^\parallel)=0$). As known, the polarization operator tensor can be expanded in a superposition of independent transverse Lorentz tensors. The number of these basic transverse tensors depends on the symmetries of the system under consideration. For example, in vacuum, where the only available tensorial structures are the four-momentum and the metric tensor, there is only one   gauge invariant structure. When a medium is under consideration (i.e. at finite temperature or finite density), since the Lorentz symmetry is broken, there is an additional gauge invariant structure that can be formed by taking into account a new four-vector, the four-velocity of the medium center of mass, $u_\mu$, \cite{Fradkin}. When a magnetic field is applied on that medium, then the structure of the polarization operator is enriched by an additional tensor, $F_{\mu\nu}$. Then, at finite density and in the presence of a magnetic field, there are nine independent gauge-invariant tensorial structures \cite{Shabad}. At strong magnetic field, when the particles are confined to the LLL, due to the fact that the transverse momentum is zero, there is a dimensional reduction leaving only the tensors $g_{\mu\nu}^\parallel, p_\mu^\parallel$ and $u^\|_\mu=(1,0)$ at our disposal. The original nine structures in \cite{Shabad} now reduce to only two
\begin{equation}
T_{\mu\nu}^{(1)}=(p^\parallel)^2g_{\mu\nu}^\parallel-p_\mu^\parallel p_\nu^\parallel,
\label{structure-1}
\end{equation}
and
\begin{equation}
T_{\mu\nu}^{(2)}=\left[u_\mu^\parallel-\frac{p_\mu^\parallel(u^\parallel\cdot p^\parallel)}{(p^\parallel)^2}\right]\left[u_\nu^\parallel-\frac{p_\nu^\parallel(u^\parallel\cdot p^\parallel)}{p^2}\right].
\label{structure-2}
\end{equation}
Moreover, one can readily check that the two tensors (\ref{structure-1}) and (\ref{structure-2}) are equivalent, which indicates that the rotated-photon polarization operator tensor, at strong magnetic field, only has one independent structure
\begin{equation}
\Pi_{\mu\nu}^\parallel(p^\parallel)=\Pi(p^\parallel,\mu,B)\left[(p^\parallel)^2g_{\mu\nu}^\parallel-p_\mu^\parallel p_\nu^\parallel\right],
\label{Cov-structure}
\end{equation}
with $\Pi(p^\parallel,\mu,B)$ being a scalar coefficient depending on the photon longitudinal momentum, baryonic chemical potential and magnetic field.

At zero temperature, the regularized components of the polarization operator in powers of the photon momentum components $p_0$ and $p_3$, up to quadratic terms are given by
\begin{equation}
{\Pi_{00}}_R=-\lim_{\Lambda\rightarrow\infty}\frac{{\tilde e}^2|{\tilde e}{\tilde B}|p_3^2}{6\pi^2}\left(\frac{1}{\Delta_0^2}+\frac{1}{\Lambda^2}\right)=-\frac{{\tilde e}^2|{\tilde e}{\tilde B}|p_3^2}{6\pi^2\Delta_0^2},
\label{Pi-00}
\end{equation}
\begin{equation}
{\Pi_{33}}_R=-\lim_{\Lambda\rightarrow\infty}\frac{{\tilde e}^2|{\tilde e}{\tilde B}|}{6\pi^2}\left[\left(3+\frac{p_0^2}{\Delta_0^2}\right)-\left(3+\frac{p_0^2}{\Lambda^2}\right)\right]=-\frac{{\tilde e}^2|{\tilde e}{\tilde B}|p_0^2}{6\pi^2\Delta_0^2}.
\label{Pi-33}
\end{equation}
and ${\Pi_{30}}_R={\Pi_{03}}_R\simeq 0$. As should be expected, the regulator $\Lambda$ introduced through the Pauli-Villars regularization scheme does not appear in the final results once we take $\Lambda \rightarrow \infty$.

Because $\Pi_{00}$ has no constant contribution in the infrared limit $p_0=0, p_3\rightarrow 0$, one immediately concludes that there is no Debye screening in the strong-field region, as it was the case at zero field in the CFL phase \cite{Wang-1, Wang-2}. This is simply because all quarks are bound within the rotated-charge neutral condensates. There is also no Meissner screening (i.e. $\Pi_{33}$ is zero in the zero-momentum limit), as it should be expected from the remnant $\tilde U(1)$ gauge symmetry. However, the condensates have electric dipole moments and could align themselves in an electric field. Hence, this should modify the dielectric constant of the medium. Since the quadratic term in the effective $\tilde U(1)$ Lagrangian is given by ${\tilde A}_\mu(-p)[D^{-1}_{\mu\nu}(p)+\Pi_{\mu\nu}(p)]{\tilde A}_\nu(p)$, with $D^{-1}$ being the bare rotated photon propagator, the effective action of the $\tilde U(1)$ field in the strong-field region is given by
\begin{equation}
S_{eff}=\int d^4x[\frac{\epsilon_\parallel}{2}{\tilde {\bf E}}_\parallel\cdot {\tilde {\bf E}}_\parallel+\frac{\epsilon_\perp}{2}{\tilde {\bf E}}_\perp\cdot {\tilde {\bf E}}_\perp-\frac{1}{2\lambda_\parallel}{\tilde {\bf H}}_\parallel\cdot {\tilde {\bf H}}_\parallel
-\frac{1}{2\lambda_\perp}{\tilde {\bf H}}_\perp\cdot {\tilde {\bf H}}_\perp],
\label{Action}
\end{equation}
where the separation between transverse and longitudinal parts is due to the $O(3) \rightarrow O(2)$ symmetry breaking produced by the strong magnetic field $\widetilde{B}$. In (\ref{Action}), $\widetilde{E}$, $\widetilde{H}$ are weak electric and magnetic field probes, respectively. In (\ref{Action}) the coefficients $\epsilon$ and $\lambda$ denote the electric permittivity and magnetic permeability of the medium respectively.

From (\ref{Pi-00})-(\ref{Pi-33}) it is straightforward that in the infrared limit the transverse and longitudinal components of the electric permittivity and magnetic permeability become
\begin{equation}
 \lambda_\perp=\lambda_\parallel\simeq 1, \quad
\epsilon_\perp=1, \quad \epsilon_\parallel=1+\chi_{MCFL}^\parallel=1+\frac{{\tilde e}^2|{\tilde e}{\tilde B}|}{6\pi^2\Delta_0^2},
\label{susceptibility}
\end{equation}
where $\chi_{MCFL}^\parallel$ is the longitudinal electric susceptibility. Notice that the longitudinal electric susceptibility is much larger than one because in the strong-magnetic-field limit ${\tilde e}{\tilde B}\gg \Delta_0^2$ \cite{SpinoneCFL}.

Although a static $\tilde U(1)$ charge cannot be completely Debye screened by the $\tilde U(1)$ neutral Cooper pairs, it can still be partially screened along the magnetic field direction because the medium is highly polarizable on that direction. This is due to the existence of Cooper pairs with opposite rotated charges $\widetilde{Q}$ that behave as electric dipoles with respect to the rotated electromagnetism of the MCFL phase. Moreover, the electric susceptibility depends on the magnetic field. When the magnetic field increases in the strong-field region, the susceptibility  becomes smaller, because the coherence length $\xi \sim 1/\Delta_0 $ decreases (i.e. $\Delta_0$ increases) with the field at a quicker rate than $\sqrt{{\tilde e}{\tilde B}}$ \cite{SpinoneCFL}, and the pair's coherence length $\xi$ plays the role of the dipole length.  Hence, with increasing magnetic field the polarization effects weaken in the strong-field region. The tuning of the electric polarization by a magnetic field is what is called in condensed matter physics the magnetoelectric effect. From (\ref{susceptibility}), we also see that at strong magnetic fields the medium turns out to be very anisotropic. The fact that the electric permittivity is only modified in the longitudinal direction is due to the confinement of the quarks to the LLL at high enough fields.

\section{Paramagnetism in Color Superconductivity}
\label{Paramagnetism}

Another nontrivial electromagnetic effect in cold-dense QCD is that an applied magnetic field can interact inside the color superconductor with the gluons, which as known, are neutral with respect to the conventional electromagnetism in vacuum.

Thus, we now analyze how the gluons are affected by an applied magnetic field in a CS state and how at sufficiently strong magnetic fields a new phase, that we call the Paramagnetic-CFL (PCFL) phase \cite{Vortex, Vortex-2}, is created. In the color superconductor some of the gluons acquire rotated electric charges. In the CFL phase the $\widetilde{Q}$-charge of the gluons in units of $\widetilde{e}$ are
\begin{equation}\label{table-2}
\begin{tabular}{cccccccc}
\hline
\textrm{$G_{\mu}^{1}$}&
\textrm{$G_{\mu}^{2}$}&
\textrm{$G_{\mu}^{3}$}&
\textrm{$G_{\mu}^{+}$}&
\textrm{$G_{\mu}^{-}$}&
\textrm{$I_{\mu}^{+}$}&
\textrm{$I_{\mu}^{-}$}&
\textrm{$\widetilde{G}_{\mu}^{8}$}\\
0 & 0 & $0$ & 1 & -1 & $1$ & $-1$ & 0\\
\hline
\end{tabular}
\end{equation}
The $\widetilde{Q}$-charged fields in (\ref{table-2}) correspond to the
combinations $G_{\mu}^{\pm}\equiv\frac{1}{\sqrt{2}}[G_{\mu}^{4}\mp
iG_{\mu}^{5}]$ and
$I_{\mu}^{\pm}\equiv\frac{1}{\sqrt{2}}[G_{\mu}^{6}\mp iG_{\mu}^{7}]$.

To investigate the effect of the applied rotated magnetic field
$\widetilde{H}$ on the charged gluons, we should start from the effective action of the charged fields $G_{\mu}^{\pm}$ (the
contribution of the other charges gluons $I_{\mu}^{\pm}$ is similar)
\begin{eqnarray}
\label{Eff-Act-3} \Gamma_{eff}& = & \int dx
\{-\frac{1}{4}(\widetilde{f}_{\mu
\nu})^{2}+G_{\mu}^{-}[(\widetilde{\Pi}_{\mu}\widetilde{\Pi}_{\mu})\delta_{\mu
\nu}-2i\widetilde{e}\widetilde{f}_{\mu \nu}\nonumber
 \\
& - & (m_{D}^{2} \delta_{\mu 0} \delta_{\nu 0}+ m_{M}^{2}
\delta_{\mu i} \delta_{\nu i})-(1-\frac{1}{\varsigma}
\widetilde{\Pi}_{\mu}\widetilde{\Pi}_{\nu})]G_{\nu}^{+}\}
\end{eqnarray}
Here, $\varsigma$ is the gauge fixing parameter, $\widetilde{\Pi}_{\mu}=\partial_{\mu}
-i\widetilde{e}\widetilde{A}_{\mu}$ is the covariant derivative in the presence of the external rotated field, $m_{D}$ and $m_{M}$ are the $G_{\mu}^{\pm}$-field Debye and Meissner masses respectively, and the field strength tensor for the rotated electromagnetic field if denoted by $\widetilde{f}_{\mu
\nu}=\partial_{\mu}\widetilde{A}_{\nu}-\partial_{\nu}\widetilde{A}_{\mu}$.
 The corresponding Debye and Meissner masses in (\ref{Eff-Act-3})
are given by \cite{Wang-1, Wang-2}
\begin{equation}
m_{D}^{2} = m_{g}^{2} \frac{21-8 \ln 2}{18},\qquad m_{M}^{2} =
m_{g}^{2} \frac{21-8 \ln 2}{54},
\end{equation}
with $m_{g}^{2}=g^2(\mu^{2}/2\pi^{2})$. We
are neglecting the correction produced by the applied field to the
gluon Meissner masses since it will be a second order effect. The effective action (\ref{Eff-Act-3}) is characteristic of a
spin-1 charged field in a magnetic field (for details see for
instance \cite{emilio, emilio-1}).

Assuming an applied magnetic field
along the third spatial direction
($\widetilde{f}^{ext}_{12}=\widetilde{H}$), we find after diagonalizing the
mass matrix of the field components ($G^{+}_{1}, G^{+}_{2}$) in (\ref{Eff-Act-3})
\begin{equation}
%\begin{eqnarray}
%\label{gapMCFL}
\left(
\begin{array}{cc}
m_{M}^{2}& i\widetilde{e}\widetilde{H} \\
- i\widetilde{e}\widetilde{H}& m_{M}^{2}
 \label{mass-matrx}
\end{array} \right) \rightarrow
\left(
\begin{array}{cc}
m_{M}^{2}+\widetilde{e}\widetilde{H}& 0 \\
0& m_{M}^{2}-\widetilde{e}\widetilde{H}
 \label{mass-matrx}
\end{array} \right),
%\end{eqnarray}
\end{equation}
with corresponding eigenvectors ($G^{+}_{1}, G^{+}_{2}$)
$\rightarrow$ ($G,iG$). We see that the lowest mass mode in (\ref{mass-matrx}) has a sort of
"Higgs mass" above the critical field
$\widetilde{e}\widetilde{H}_{C}= m_{M}^2$, indicating the setup of an instability for the
$G$-field. This phenomenon is the well known "zero-mode
problem" found in the presence of a magnetic field for Yang-Mills
fields \cite{zero-mode-1, zero-mode-2}, for the $W^{\pm}_{\mu}$ bosons in the
electroweak theory \cite{Olesen, Olesen-1,Skalozub}, and even for
higher-spin fields in the context of string theories
\cite{porrati-1, porrati-2} and it is due to the presence of the gluon anomalous magnetic moment term $2i\widetilde{e}\widetilde{f}_{\mu
\nu}G_{\mu}^{-}G_{\nu}^{+} $ in (\ref{Eff-Act-3}).
Thus, to remove
the instability it is needed the restructuring of the ground
state through the condensate of the field bearing the tachyonic mode (i.e. the $G$-field).

It is worth to call attention that the gluon condensate under consideration is not the
only charged spin-one condensate generated in a theory with a large fermion density.
As known \cite{Linde-1, Linde-2, Linde-3}, a spin-one condensate of W-bosons can be originated at sufficiently high
fermion density in the context of the electroweak theory at zero magnetic field. However,
the physical implications of the gluon condensate induced by the magnetic field in the
CS are fundamentally different from those associated to the homogeneous W-boson
condensate of the dense electroweak theory \cite{Linde-1, Linde-2, Linde-3}. One of the main physical differences is that the homogeneous W condensate, being electrically charged, so to compensate the excess of charge due to the finite density of electrons \cite{Linde-1, Linde-2, Linde-3}, breaks the electromagnetic $U(1)$ group producing a conventional superconducting state \cite{W-condensate}; while the inhomogeneous gluon condensate in CS is formed with gluons of both charges, so keeping the condensate state neutral.

To find the $G$-field condensate and the induced magnetic field
$\widetilde{\textbf{B}}=\nabla\times\widetilde{\textbf{A}}$, with
$\widetilde{\textbf{A}}$ being the total rotated electromagnetic
potential in the condensed phase in the presence of the external
field $\widetilde{H}$, we should start from the Gibbs free energy
density $\mathcal{G}=\mathcal{F}-\widetilde{H}\widetilde{B}$, since
it depends on both $\widetilde{B}$ and $\widetilde{H}$
($\mathcal{F}$ is the system free energy density). Since
specializing $\widetilde{H}$ in the third direction the instability
develops in the $(x,y)$-plane, we make the ansatz for the condensed field $\overline{G}=\overline{G}(x,y)$.
Starting from (\ref{Eff-Act-3}) in the Feynman gauge $\varsigma=1$,
which in terms of the condensed field $\overline{G}$ implies
$(\widetilde{\Pi}_{1}+i\widetilde{\Pi}_{2})\overline{G}=0$, we have that the
Gibbs free energy in the condensed phase is
\begin{equation}
\label{Gibbs} \mathcal{G}_{c} =\mathcal{F}_{n0} +\widetilde{\Pi}^{2}
\overline{G}^{2}-2(\widetilde{e}\widetilde{B}-m_{M}^{2})\overline{G}^{2}+2g^{2}\overline{G}^{4}+\frac{1}{2}\widetilde{B}^{2}-\widetilde{H}\widetilde{B}.
\end{equation}
where $\mathcal{F}_{n0}$ is the system free energy in the normal
phase ($\overline{G}=0$) at zero magnetic field.

The minimum equations for the fields $\overline{G}$ and
$\widetilde{B}$ are respectively obtained from (\ref{Gibbs}) as
\begin{equation}
\label{EqG} \widetilde{\Pi}^{2}
\overline{G}+2(m_{M}^{2}-\widetilde{e}\widetilde{B})\overline{G}+8g^{2}\overline{G}^{2}\overline{G}=0,
\end{equation}

\begin{equation}
\label{EqB} 2\widetilde{e} \overline{G}^{2}-\widetilde{B}+\widetilde{H}=0
\end{equation}
Identifying $\overline{G}$ with the complex order parameter, Eqs.
(\ref{EqG})-(\ref{EqB}) become analogous to the Ginzburg-Landau
equations for a conventional superconductor except by the negative sign in front of the $\widetilde{B}$ field in Eq. (\ref{EqG}) and the positive sign in the first term of the LHS of Eq. (\ref{EqB}) \cite{Vortex}. The fact that those signs turn the opposite of those appearing in conventional superconductivity is due to the different nature of the condensates in both cases. While in conventional superconductivity the Cooper pair is a spin-zero condensate, here we have a condensate formed by spin-one charged particles interacting through their anomalous magnetic moment with the magnetic field (i.e. the term $2i\widetilde{e}\widetilde{f}_{\mu
\nu}G_{\mu}^{-}G_{\nu}^{+} $ in (\ref{Eff-Act-3})).

Notice that because of the different sign in
the first term of (\ref{EqB}), the resultant field $\widetilde{B}$
is stronger than the applied field $\widetilde{H}$, contrary to what occurs in
conventional superconductivity. Thus, when a
gluon condensate develops, the magnetic field will be antiscreened
and the color superconductor will behave as a paramagnet. The
antiscreening of a magnetic field has been also found in the context
of the electroweak theory for magnetic fields $H \geq
M_{W}^{2}/e\sim 10^{24} G$ \cite{Olesen}. Just as in the electroweak
case, the antiscreening in the color superconductor is a direct
consequence of the asymptotic freedom of the underlying theory
\cite{Olesen}.

Therefore, the magnetic field in the new
phase is boosted to a higher
value, which depends on the modulus of the $\overline{G}$-condensate. That is why the phase attained at $\widetilde{H}\geq \widetilde{H}_c$ is called paramagnetic CFL (PCFL) \cite{ Vortex,phases}. It should be pointed out that at the scale of baryon densities
typical of neutron-star cores ($\mu \simeq 400 MeV$, $g(\mu)\simeq
3$) the charged gluons magnetic mass in the CFL phase is $m_{M}^{2}
\simeq 16\times 10^{-3} GeV^{2}$. This implies a critical magnetic
field of order $\widetilde{H}_{c}\simeq 0.7\times 10^{17} G$. Although it is a significant high value, it is in the expected range
for the neutron star interiors with cold-dense quark matter \cite{H-Estimate, EoS-H}. Let us underline that in our analysis
we considered asymptotic densities where quark masses can be
neglected. At lower densities where the Meissner masses of the charged gluons become smaller, the field values needed to develop the magnetic instability will be smaller.

To find the structure of the gluon condensate we should solve the non-linear differential equation (\ref{EqG}). However, to get an analytic solution we can consider the approximation where  $\widetilde{H}\approx \widetilde{H}_c=m_M^2$ and consequently $\mid \overline{G}\mid \approx 0$. In this approximation, Eq. (\ref{EqG}) can be linearized as
\begin{equation}
\label{EqVortex} [\partial_{j}^{2}-\frac{4\pi
i}{\widetilde{\Phi}_{0}}\widetilde{B}x\partial_{y}-4\pi^{2}\frac{\widetilde{B}^{2}}{\widetilde{\Phi}_{0}^{2}}x^{2}-\frac{1}{\xi^{2}}]\overline{G}=0,
\qquad j=x,y
\end{equation}
where we fixed the gauge condition
$\widetilde{A}_{2}=\widetilde{B}x_{1}$, and introduced the notations
$\widetilde{\Phi}_{0}=2\pi/\widetilde{e}$, and
$\xi^{2}=\frac{1}{2}(\widetilde{e}\widetilde{B}-m_{M}^{2})^{-1}$.

Eq. (\ref{EqVortex}) is formally similar to the Abrikosov's equation in type-II conventional superconductivity \cite{Abrikosov, Abrikosov-1, Abrikosov-2}, with $\xi$ playing the role of the coherence length and $\widetilde{\Phi}_{0}$ of the flux quantum per vortex cell. Then, following the Abrikosov's approach, a solution of Eq. (\ref{EqVortex}) can be found as
\begin{equation}
\label{Vortex-solution} \overline{G}(x,y)=\frac{1}{\sqrt{2}\widetilde{e}\xi}e^{-\frac{x^2}{e\xi^2}}\vartheta_3(u/\tau),
\end{equation}
with $\vartheta_3(u/\tau)$ being the elliptic theta function with arguments
\begin{equation}
\label{arguments} u=-i\pi b(\frac{x}{\xi^2}+\frac{y}{b^2}), \qquad \tau=-i\pi\frac{b^2}{\xi^2}
\end{equation}
In (\ref{arguments}) the parameter $b$ is the periodic length in the y-direction ($b=\Delta y$). The double periodicity of the elliptic theta function also implies that there is a periodicity in the x-direction given by $\Delta x=\widetilde{\Phi}_{0}/b\widetilde{H}_{c}$. Therefore, the magnetic flux through each
periodicity cell ($\Delta x \Delta y$) in the vortex lattice is quantized $\label{Flux}
\widetilde{H}_c \Delta x \Delta y=\widetilde{\Phi}_{0}$, with
$\widetilde{\Phi}_{0}$ being the flux quantum per unit vortex cell.
In this semi-qualitative analysis we considered the Abrikosov's
ansatz of a rectangular lattice, but the lattice configuration should be carefully
determined from a minimal energy analysis. For the rectangular
lattice, we see that the area of the unit cell is $A=\Delta x \Delta
y=\widetilde{\Phi}_{0} /\widetilde{H}_c$, so decreasing with
$\widetilde{H}$.

In conclusion,  to remove the instability created by an external uniform
magnetic field in the z-direction, a periodic arrangement of  vortices of charged gluon condensates is generated in the $(x,y)$-plane. The currents in the $(x,y)$-plane created by these vortices increase the magnitude of the net magnetic field in the direction of the original field, but since the magnitude of the resultant field varies in the $(x,y)$-plane, the vortex condensate leads to a net inhomogeneous magnetic field. Therefore, the presence of a supercritical magnetic field leads to the formation of a fluxoid along
the $z$-direction and the appearance of a nontrivial topology on the
perpendicular plane. From (\ref{EqB}) we see that the resultant magnetic
field can go from a minimum value $\widetilde{H}$ to a maximum at
the core of the fluxoid that depends on the amplitude of the gluon
condensate determined by the mismatch between the applied field and
the gluon Meissner mass.

Summarizing, at low $\widetilde{H}$ field, the CFL phase behaves as an
insulator, and the $\widetilde{H}$ field just penetrates through it without any change of strength.
At sufficiently high field $\widetilde{e}\widetilde{H}\sim m_M^2$, the condensation of $G^{\pm}$
is triggered inducing the formation of a lattice of magnetic flux
tubes that breaks the translational and remaining rotational
symmetries, creating the so called paramagnetic phase. We stress that contrary to the situation in conventional type-II
superconductivity, where the applied field only penetrates through the
flux tubes and with a smaller strength, the vortex state in the color superconductor has the peculiarity that outside the flux tube the
applied field $\widetilde{H}$ totally penetrates the sample, while
inside the tubes the magnetic field becomes larger than
$\widetilde{H}$ (this is the origin of the paramagnetic behavior of this CS phase). This effect provides an internal mechanism to increase the magnetic field of a compact star with a CS core.

\section{Magnetic Phases in CFL Matter}
\label{Phases}

From the discussions in the previous sections it is clear that in the three-flavor color superconductor at
very high densities an increasing magnetic field produces a
crossover from CFL to MCFL first, and then a phase transition
from MCFL to PCFL. 

During the crossover, no symmetry
breaking occurs, since in principle once a magnetic field is present
the symmetry is already that of the MCFL.  At very weak magnetic fields, the color superconducting state is
practically described by the CFL phase, because the charged mesons corresponding to the Goldstone modes,
although massive, are so light that they cannot decay in pairs of
quark-antiquark. When the field strength is of the order of the
quarks' energy gap $\Delta_{CFL}$, the charged mesons become heavy enough to
decouple and the low-energy physics is indeed that of the MCFL
phase, where five neutral massless mesons drive the low-energy
behavior.

Going from MCFL to
PCFL is, on the other hand, a real phase transition \cite{Vortex, Vortex-2},
as the translational symmetry, as well as the remaining
rotational symmetry in the plane perpendicular to the applied
magnetic field are broken by the vortex state. This phase transition is driven by fields whose strengths are comparable to the magnetic masses $m_M$ of the charged
gluons, so creating a chromomagnetic instability that leads to the formation of a vortex state and the antiscreening of
the magnetic field \cite{Vortex, Vortex-2}.

This magnetic instability is characteristic of systems of charged bosons with higher spins ($s\geq1$). Taking into account that at zero momentum the energy spectrum in a magnetic field H of a charged boson of spin s,
charge e, gyromagnetic ratio g, and mass m is
\begin{equation}\label{spin-1}
E^2_n = (2n + 1)eH - geH \cdot s + m^2,
\end{equation}
it is evident that for spin-one particles, for which g = 2, the energy becomes imaginary, i.e. $E^2 < 0$,  if the field satisfies $H > H_{cr} = m^2/e$), implying that when the field surpasses
the critical value $H_{cr}$, one of the modes of the charged gauge field becomes tachyonic inducing the vortex formation.

Within a NJL model, for fields comparable to the baryon chemical potential, the ground state is that of the MCFL phase with sizable values of the three condensates $\Delta_B$, $\Delta_M$,  and $\Delta$.  However, once the gluon effects are taken into account, the PCFL vortex state generated at lower fields is unavoidable and the picture becomes much more complicated due to the inhomogeneities of the gluon condensate and net magnetic field. From a physical point of view,  it is natural to expect that in this situation the three fermion gaps  will remain, because their physical origin, is still the same. That is, an inhomogeneous magnetic field will also distinguish between pairs of opposite charged quarks and pairs of neutral quarks, and those of opposite charged quarks will still have a magnetic moment contributing to the condensate $\Delta_M$. However, all these condensates should become inhomogeneous in the $(x,y)$-plane.  

\section{Equation of State of the MCFL Phase}
\label{EoS}

At present, some of the best-known characteristics of stellar
objects are their masses and radii. The relation between the mass and the
radius of a star is determined by the equation of state (EoS) of the inner phase
of the matter in the star. If one can identify some features  connecting
the star's internal state (nuclear,
strange, color superconducting, etc.) to its mass/radius relation, one
would have an observational tool to discriminate among the actual
realization of different star inner phases in nature. From previous
theoretical studies \cite{MassRadius-1, Alford03, MassRadius-2, MassRadius-3, MassRadius-4, MassRadius-5, MassRadius-6,German, MassRadius-7, MassRadius-9, MassRadius-8, MassRadius-10, MassRadius-11}
the mass-radius relationship predicted for neutron stars with
different quark-matter phases (CS or unpaired) at the core are very
similar to those having hadronic phases, at least for the observed mass/radius range.
As a consequence, it
is very difficult to find a clear observational signature that
can distinguish among them. Nevertheless, an important ingredient
was ignored in these studies: the magnetic field, which in some compact
stars could reach very high values in the inner regions.

As pointed out in \cite{H-Estimate},
a strong magnetic field can create a significant
anisotropy in the
longitudinal and transverse pressures.
One would expect then, that the EoS, and consequently,
the mass-radius ratio, become affected by sufficiently strong core fields. Given
that we are beginning to obtain real observational constraints on the
EoS of neutron stars \cite{Alford2010},
it is important to investigate the EoS in the presence of a magnetic
field for different inner star phases to be able to discard those
that do not agree with observations.

In order to understand the relevance of the magnetic field to
tell apart neutron stars from stars with paired quark matter,
it is convenient to recall that when the pressure
exerted by the central matter density of neutron stars (which is about
$200-600 MeV/fm^3$) is contrasted with that exerted by an electromagnetic
field, the field strength needed for these two contributions to be
of comparable order results of order $\sim 10^{18} G$ \cite{Broderick}.
It is worth to notice that even these very strong fields are not enough to produce quantum effects like the Landau quantization of the protons, because these effects only show up when the particles's cyclotron energy
$ehB/mc$ becomes comparable to its rest energy $mc^{2}$,
which for protons means a field $\sim 10^{20} G$.

However, for stars with paired quark matter, the situation
is rather different. Naively, one might think that comparable matter and field pressures
in this case would occur only at
much larger fields, since the quark matter can only exist at even larger densities
to ensure deconfinement. In reality, though, the situation is more subtle. As argued
in \cite{Alford03}, the leading term in the matter pressure coming from the contribution
of the particles in the Fermi sea, $\sim\mu^{4}$, could be (almost) canceled out by the negative
pressure of the bag constant and in such a case, the next-to-leading term would play a
more relevant role than initially expected. Consequently, the magnetic pressure might
only need to be of the order of that produced by the particles close to the Fermi surface,
which becomes the next-to-leading contribution, $\sim\mu^{2}\Delta^{2}$, with $\Delta$ the
superconducting gap and $\mu$ the baryonic chemical
potential. For typical values of these parameters in paired quark matter one obtains a
field strength $\sim 10^{18} G$. Moreover, the magnetic field can affect the pressure
in a less obvious way too, since as shown in \cite{MCFL-1, MCFL-2, MCFL-3},
it modifies the structure and magnitude of the superconductor's gap, an effect that, as found in \cite{Vortex, Vortex-2}, starts to become relevant already at fields of order $10^{16} G$
and leads to de Haas van-Alphen oscillations of the gap magnitude \cite{MCFLoscillation,MCFLoscillation-1}.
It is therefore quite plausible that the effects of moderately
strong magnetic fields in the EoS of compact
stars with color superconducting matter will be more noticeable than in
stars made up only of nucleons, where quantum effects starts to be
significant for field four orders of magnitude larger. This is why an evaluation of the EoS in magnetized
quark phases is necessary and relevant.

In \cite{EoS-H}, a self-consistent analysis of the EoS of MCFL matter, was performed taking into consideration the solution of the gap equations and the anisotropy of the pressures in a magnetic field. In that study a uniform and constant magnetic field was assumed. The reliability of this assumption for neutron stars, where the magnetic field strength is expected to vary from the core to the surface in several orders, is based on the fact that the scale of the field variation in the stellar medium is much larger than the microscopic magnetic scale for both weak and strong magnetic fields \cite{Broderick}. Hence, when investigating the field effects in the EoS, it is consistent to take a magnetic field that is locally constant and uniform. This is the reason why such an approximation has been systematically used in all the previous works on magnetized nuclear \cite{MNS-1, MNS-2, Broderick, MNS-3, MNS-4, MNS-5, MNS-6, MNS-7, MNS-8, MNS-9, MNS-10, MNS-11, MNS-12, MNS-13, MNS-14, MNS-15} and quark matter \cite{MQS-1, MQS-2, MQS-3, MQS-4, MQS-5, MQS-6}.

\subsection{Covariant Structure of the Energy-Momentum Tensor in a Magnetized System}
\label{Covariant-T}

In the reference frame comoving with the many-particle system, the system normal stresses (pressures) can be obtained from the diagonal spatial components of the average energy-momentum tensor $\langle \tau^{ii} \rangle$; the system energy, from its zeroth diagonal component $\langle \tau^{00} \rangle$; and the shear stresses (which are absent for the case of a uniform magnetic field) from the off-diagonal spatial components $\langle \tau^{ij} \rangle$ \cite {Landau-Lifshitz}. Then, to find the energy density and pressures of the dense magnetized system we need to calculate the quantum-statistical averages of the corresponding components of the energy-momentum tensor of the fermion system in the presence of a magnetic field.

These calculations were carried out long time ago in Ref. \cite{Canuto}, using a QFT second-quantization approach. There, a quantum-mechanical average of the energy-momentum tensor  in the eigen-states of the Dirac equation in the presence of the uniform magnetic field was first performed to get the corresponding quantum operator in the occupation-number space. The macroscopic stress-energy tensor was then found by averaging its quantum operator in the statistical ensemble using the many-particle density matrix.  Similar calculations were performed in Ref. \cite{H-Estimate},  but using a functional-method approach that makes it easier to recognize the thermodynamical quantities entering in the final results. An advantage of the procedure followed in \cite{H-Estimate}, as compared with that of \cite{Canuto}, is that it does not assume that the fermion fields entering in the definitions of the energy and pressures satisfy the classical equation of motions (i.e. the Dirac equations for $\psi$ and $\overline{\psi}$), but the functional integrals integrate in all field configurations. Hence,  the terms depending on the Lagrangian density $\cal{L}_\psi$ in $\tau_{\mu\nu}$ were kept, while in Ref. \cite{Canuto} the condition $\cal{L}_\psi$$=0$ was considered as a constraint.

Let us then consider a hot and dense system of fermions in a constant and uniform magnetic field $B$. At this point it is convenient to introduce the covariant decomposition for the energy momentum tensor of the whole system containing the matter and field contributions. In order to accomplish this goal, we define the system thermodynamic potential as the sum of the matter, $\Omega_f$, and field, $B^2/2$, contributions
\begin{equation}\label{System-Omega}
\Omega=\Omega_f+\frac{B^2}{2}
\end{equation}

Taking into account the symmetries of the magnetized dense system, we can write the statistical average of the energy-momentum tensor as a combination of all the available independent structures
\begin{equation}\label{System-T}
\frac{1}{\beta V}\langle \widetilde{\tau}^{\mu\nu}\rangle=\Omega \eta^{\mu\nu}+(\mu N+TS)u^{\mu}u^{\nu}+BM\eta^{\mu\nu}_\perp,
\end{equation}
where $N=-(\partial\Omega/\partial \mu)$ is the particle number density, $S=-(\partial\Omega/\partial T)$ is the system entropy, $M=-(\partial\Omega/\partial B)$ is the system magnetization and $\eta_{\perp}^{\mu \nu}=\widehat{F}^{\mu \rho}\widehat{F}_{\rho}^\nu$ (where $\widehat{F}^{\mu \rho}=F^{\mu \rho} /B$ denotes the normalized electromagnetic strength tensor).

To understand the origin of the covariant decomposition (\ref{System-T}), notice that as a consequence of the breaking of the rotational symmetry $O(3)$ produced by the external magnetic field, the Minkowskian metric splits in transverse $\eta_{\perp}^{\mu \nu}$ and longitudinal $\eta_{\|}^{\mu \nu}=\eta^{\mu \nu}-\widehat{F}^{\mu \rho}\widehat{F}_{\rho}^\nu$ structures.
Considering the quantum field limit with no magnetic field, i.e. when $T=\mu=B=0$, the only term different from zero is the first one in the RHS of (\ref{System-T}). In that case the system has Lorentz symmetry and the energy density, $\varepsilon$, and pressure, $p$, are given by $\varepsilon=-p=\Omega_f$. If temperature and/or density are switched on, then the Lorentz symmetry is broken specializing a particular reference frame comoving with the medium center of mass and having four velocity $u_\mu=(1,\overrightarrow{0})$. This is reflected in the second term of the RHS of (\ref{System-T}). In this case, at $T=0$ for instance, $\varepsilon=\Omega_f+\mu N$ and $p=-\Omega_f$. Finally, when there is an external uniform magnetic field acting on the system, the additional symmetry breaking $O(3)\rightarrow O(2)$ takes place, and $\langle \widetilde{\tau}^{\mu\nu}\rangle$ get an anisotropy reflected in the appearance of the transverse metric structure $\eta^{\mu\nu}_\perp$ in (\ref{System-T}). At $T=0$ we then have

\begin{equation} \label{EOS-E-1}
\epsilon=\Omega_{f}-\mu \frac{\partial \Omega_{f}}{\partial \mu}+\frac{B^2}{2},
\end{equation}
\begin{equation}\label{EOS-P-1}
p^\|=-\Omega_{f}-\frac{B^2}{2}, \nonumber \quad
 p^\bot=-\Omega_{f}+H \frac{\partial\Omega_{f}}{\partial B}+\frac{B^2}{2}
\end{equation}

See Ref. \cite{H-Estimate} for detailed derivations of
the formulas for the pressures and energy density in a magnetic field.

\subsection{MCFL Thermodynamic Potential}
\label{Therm-Pot}

Let us turn our attention now to densities large enough for the fermion system to be in the MCFL phase. Our ultimate goal is to find the EoS of this superconducting phase. To find the density and pressure of this phase, we first need, as seen from  (\ref{EOS-E-1})-(\ref{EOS-P-1}), to obtain the contribution of the quarks to the thermodynamic potential. We can express the MCFL thermodynamic potential as the sum of the contributions
coming from charged ($\Omega_C$) and neutral ($\Omega_N$) quarks \cite{EoS-H}
\begin{equation}
\Omega_{MCFL} =\Omega_C+\Omega_N
\label{OmegaMCFL}
\end{equation}
\\
with

\begin{equation}\label{C}
\Omega_{C} =-\frac{\widetilde{e}\widetilde{B}}{4\pi^2}\sum_{n=0}^\infty
(1-\frac{\delta_{n0}}{2})\int_0^\infty dp_3 e^{-(p_3^2+2\widetilde{e}\widetilde{B}n)/
\Lambda^2}[8|\varepsilon^{(c)}|+8|\overline{\varepsilon}^{(c)}|],
\end{equation}

\begin{eqnarray}\label{N}
\Omega_{N} =-\frac{1}{4\pi^2}\int_0^\infty dp p^2 e^{-p^2/\Lambda^2}[6|
\varepsilon^{(0)}|+6|\overline{\varepsilon}^{(0)}|]\qquad\qquad\nonumber
\\
-\frac{1}{4\pi^2}\int_0^\infty
dp p^2 e^{-p^2/\Lambda^2}\sum_{j=1}^2[2|\varepsilon_j^{(0)}|+2|\overline{\varepsilon}_j^{(0)}|]+
\frac{\Delta^2}{G}+\frac{2\Delta^2_B}{G},
\end{eqnarray}

and
\begin{eqnarray}
\varepsilon^{(c)}=\pm \sqrt{(\sqrt{p_3^2+2\widetilde{e}\widetilde{B}n}-\mu)^2+\Delta_B^2},\nonumber
\\
\overline{\varepsilon}^{(c)}=\pm \sqrt{(\sqrt{p_3^2+2\widetilde{e}\widetilde{B}n}+\mu)^2+\Delta_B^2},
\end{eqnarray}

\begin{eqnarray}
\varepsilon^{(0)}=\pm \sqrt{(p-\mu)^2+\Delta^2},\qquad \overline{\varepsilon}^{(0)}=\pm \sqrt{(p+\mu)^2+\Delta^2},\nonumber
\\
\varepsilon_1^{(0)}=\pm \sqrt{(p-\mu)^2+\Delta_a^2},\qquad \overline{\varepsilon}_1^{(0)}=\pm
\sqrt{(p+\mu)^2+\Delta_a^2},\nonumber
\\
\varepsilon_2^{(0)}=\pm \sqrt{(p-\mu)^2+\Delta_b^2},\qquad \overline{\varepsilon}_2^{(0)}=\pm
\sqrt{(p+\mu)^2+\Delta_b^2},
\end{eqnarray}
being the dispersion relations of the charged $(c)$ and neutral $(0)$ quarks. In the above we used the notation
\begin{equation}
\Delta_{a/b}^2=\frac{1}{4}(\Delta\pm\sqrt{\Delta^2+8\Delta_B^2})^2
\end{equation}
The MCFL gaps $\Delta$ and $\Delta_{B}$ were introduced in Section 4. In the integrals (\ref{C}) and (\ref{N}), $\Lambda$-dependent smooth cutoffs for the NJL model are used.

The effects of confinement can be incorporated by adding a bag constant $\cal{B}$ to
$\Omega_{MCFL}$. Besides the bag constant and the quark contributions, the thermodynamic potential of the system also includes the pure Maxwell
contribution, $\widetilde{B}^2/2$  \cite{H-Estimate}.
Hence, the thermodynamic potential of the MCFL phase is given by
\begin{equation}\label{Omega-H}
 \Omega_{B}=\Omega_{MCFL}+{\cal{B}}+\frac{\widetilde{B}^2}{2},
\end{equation}

The gaps $\Delta$, and $\Delta_B$ have to
be found from their respective gap equations
\begin{equation}
\frac{\partial \Omega_{MCFL}}{\partial \Delta}=0,\qquad\qquad \frac{\partial \Omega_{MCFL}}{\partial \Delta_B}=0.
\label{Delta-MCFL}
\end{equation}

\subsection{EoS in a Magnetic Field}
\label{EoS-MCFL}

The pressure and energy density of the MCFL phase are given by

\begin{equation} \label{EOS-E-MCFL}
\epsilon_{MCFL}=\Omega_{B}-\mu \frac{\partial \Omega_{B}}{\partial \mu},
\end{equation}
\begin{equation}\label{EOS-P-MCFL}
p^\|_{MCFL}=-\Omega_{B}, \nonumber \quad
 p^\bot_{MCFL}=-\Omega_{B}+\widetilde{B} \frac{\partial\Omega_{B}}{\partial \widetilde{B}}
\end{equation}

Note the splitting between parallel $p^\|_{MCFL}$ (i.e.
along the field) and transverse $p^\bot_{MCFL}$ (i.e.
perpendicular to the field) pressures due to the magnetic field.

\begin{figure}
\begin{center}
\includegraphics[width=0.47\textwidth]{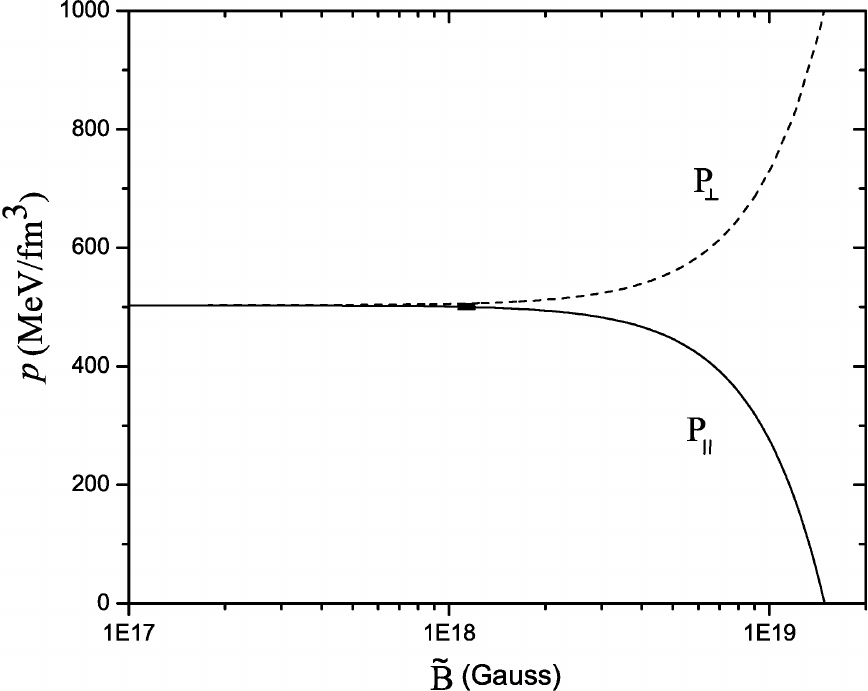}
\caption{\footnotesize Parallel and perpendicular pressures as a
function of the magnetic field intensity for
 $\mu=500 MeV$ and bag constant ${\cal{B}}=58 MeV/fm^3$.} \label{Pressures}
\end{center}
\end{figure}

The magnetic field dependencies of the parallel and transverse
pressures in (\ref{EOS-P-MCFL}) were studied in Ref. \cite{EoS-H}, and are plotted in Fig. \ref{Pressures}. Similarly
to what occurs in the case of a magnetized uncoupled fermion system at finite
density \cite{H-Estimate}, the transverse pressure in the MCFL phase increases with the field,
while the parallel pressure decreases and reaches a zero value at
field strength of order $\geq 10^{19}$ G for the density under
consideration ($\mu=500$ MeV). We see from Fig. \ref{Pressures} that $\Omega_H$ and
$\partial \Omega_B/\partial \widetilde{B}$ do not exhibit the Hass-van Alphen
oscillations as happens with other physical quantities in the presence of a magnetic
field \cite{Klimenko, Klimenko-1,SpinoneCFL,MCFLoscillation,MCFLoscillation-1}. This is due to the high contribution of the
pure Maxwell term in $\Omega_B$ and $\partial \Omega_B/\partial \widetilde{B}$,
which makes the oscillations of the matter part negligible in comparison.

\begin{figure}
\begin{center}
\includegraphics[width=0.44\textwidth]{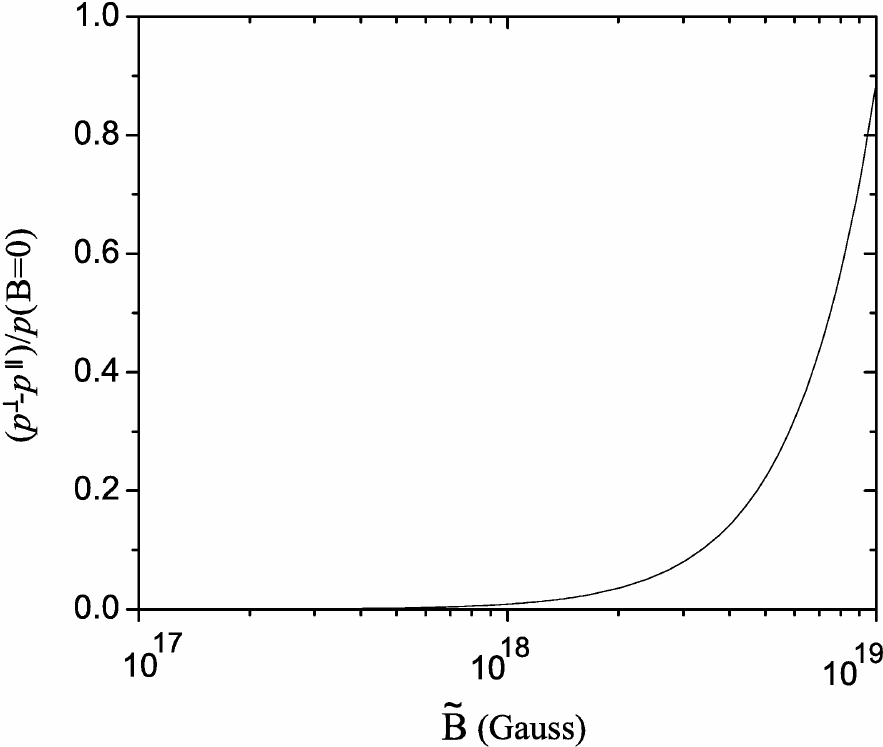}
\caption{\footnotesize Splitting of the parallel and perpendicular
pressures, normalized to the zero value pressure ($p(\widetilde{B}=0)$), as a
function of the magnetic field intensity for $\mu=500$ MeV and
${\cal{B}}=58 MeV/fm^3$.} \label{Split}
\end{center}
\end{figure}

The splitting between parallel and perpendicular pressures, shown
in the vertical axis of Fig.\ref{Split}, grows with the magnetic
field strength. Comparing the found splitting with the pressure of the (isotropic) CFL
phase, we can address how important this effect is for the
EoS. Notice that for $3 \times 10^{18}$ G the
pressures splitting is $\sim 10\%$ of their isotropic value at
zero field (i.e. the one corresponding to the CFL phase).

\begin{figure}
\begin{center}
\includegraphics[width=0.47\textwidth]{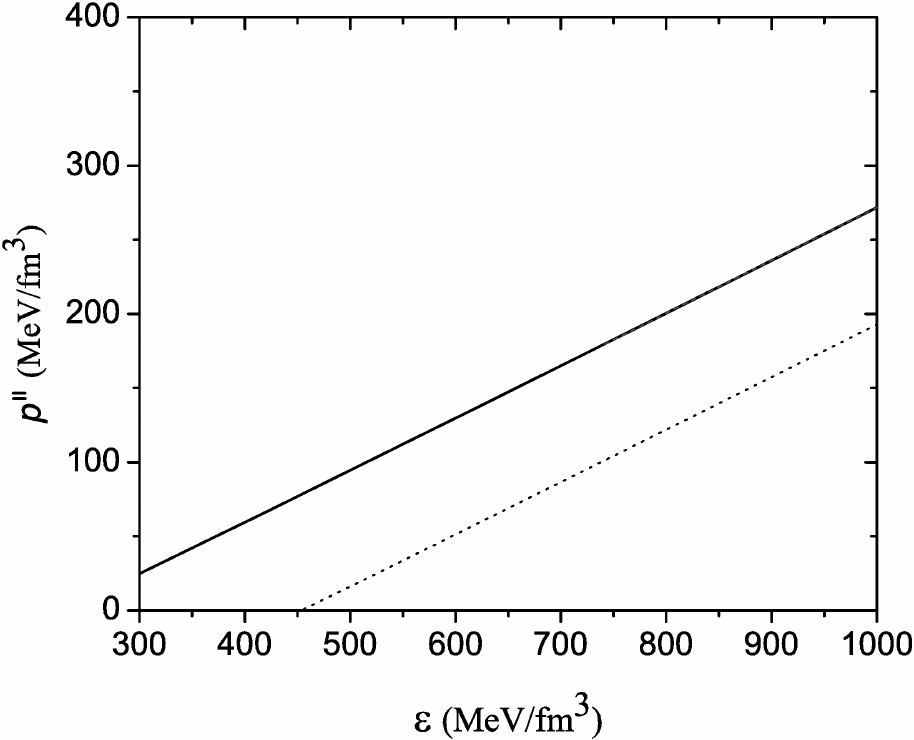}
\includegraphics[width=0.47\textwidth]{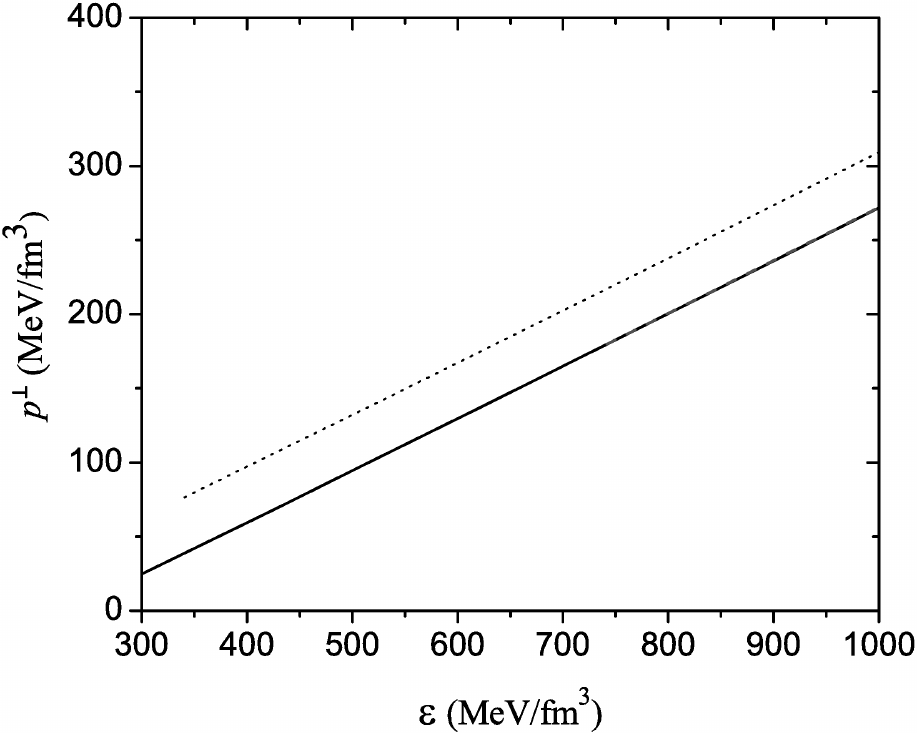}
\caption{\footnotesize Equation of state for MCFL matter
considering parallel (right panel) and perpendicular (left panel)
pressures for different values of $\widetilde{B}$: zero field (solid line),
$10^{17} G$ (dashed line) and $5 \times 10^{18} G$ (dotted line).
Note that the low value of $\widetilde{B}=10^{17} G$ is not distinguishable in
the plots, being merged with the zero-field curve. The value of
the bag constant was fixed to ${\cal{B}}=58 MeV/fm^3$.} \label{EOS}
\end{center}
\end{figure}

In the graphical representation of the EoS in Fig. \ref{EOS} the
highly anisotropic behavior of the magnetized medium is explicitly
shown. While the magnetic-field effect is significant for the
$\epsilon-p^\|$ relationship at $\widetilde{B}\sim 10^{18}$ G, with a
shift in the energy density with respect to the zero-field value of
$\sim 200$ MeV/fm$^3$ for the same pressure, the field effect in the
$\epsilon-p^\bot$ relationship is smaller for the same
range of field values.

%\begin{figure}
%\begin{center}
%\includegraphics[width=0.47\textwidth]{EnergiaXnA.eps}
%\caption{\footnotesize Energy per baryon number as a function of
%the baryonic density of MCFL matter for different values of the
%magnetic field, labeled as in Fig.3. We see that increasing the magnetic field increases the energy per baryon, thus
%making the matter less stable.} \label{EnergyA}
%\end{center}
%\end{figure}

The most important application of the EoS is to construct stellar models for compact stars composed of quark matter.
This goal can be archived by using the relativistic
equations of stellar structure, that is, the well known
Tolman-Oppenheimer-Volkoff (TOV) and mass continuity equations.

\begin{eqnarray}
\frac{dm}{dr}&=&4\pi r^2\epsilon \label{TOV1}\\
\frac{dP}{dr}&=&-\frac{\epsilon m}{r^2}\Big(1+\frac{P}{\epsilon}\Big)\Big
(1+\frac{4\pi r^3P}{m}\Big)\Big(1-\frac{2m}{r}\Big)^{-1}\label{TOV2}
\end{eqnarray}
written in natural units, $c = G = 1$. However, it is clear that this set of differential equations apply only to isotropic EoS for systems with spherical symmetry.

If the magnetic field in
the MCFL phase is high enough for the anisotropy in the pressure to be significant (i.e., expressed in terms of the pressure splitting to be
$(\Delta p/p_{CFL})\sim (\widetilde{B}^2/\mu^2\Delta^2)\sim {\cal{O}}(1)$)
the spherically symmetric TOV equations become inappropriate,
because the deviations lead to significant
differences with respect of realistic axi-symmetric models, yet to
be constructed \cite{EoS-H}.

\section{ Astrophysical Implications}
\label{Astrophysics}

As we have stressed, an important characteristic of neutron stars is that they typically possess very strong magnetic fields. Unveiling the interconnection between the star's magnetic field and the dense phases is important to understand the interplay between QCD and neutron star phenomenology. As discussed above, in recent years much interesting work has been done on the properties of the different nuclear phases that can be reached in dense astrophysical objects in the presence of strong magnetic fields. An important new step in this context would be to consider the consequences for the star's phenomenology of the possible new phases. Although at this point we do not know yet the quantitative details of the potential consequences, no doubt exploring them will shed new light on the important question of how we can infer the presence of a color-superconducting core from astronomical observations, and whether such observations can distinguish among different color superconducting phases. In what follow we discuss some of these related tasks.

\subsection{Low-Energy Physics}
\label{subsec:1}

The main challenge in determining the phases of matter inside a neutron star is to provide observables signatures of the presence of those phases. Currently there are many proposals to connect observations to the inner phases of the star. We want to discuss in general terms here, those connected to transport properties as conductivities, viscosities, etc. As known, these transport properties are determined by the low-energy spectrum of the phase, that is, by the lowest-energy modes as Goldstone bosons and gapless quark excitations, which as already shown, can be affected by the presence of a sufficiently high magnetic field. Let's briefly mention some examples of transport properties and how they could affect the star's observables.

\underline{Viscosity}. The viscosity of the interior of a star can be probed by observing how a rapidly spinning neutron star slows down. If the star slows down very quickly this indicates that it is unstable with respect to bulk flows (r-modes) that transfer the star angular momentum into gravitational radiation. But this can only occur if damping is sufficiently small. Based upon these arguments the possibility of pure CFL-quark-matter pulsars has been ruled out [51] since in the CFL phase viscous damping is negligible \cite{Manuel-Dobado}. If the intermediate-density CS phase happens to have a large viscosity, it will not be restricted by r-modes arguments tough.

As we showed in  \cite{MCFL}, in a three-flavor theory the spectrum of the NG bosons of the CS is affected by the restructuring of the gap produced by the magnetic field. As a consequence, instead of the 9 Goldstone bosons that exist in the CFL phase, in the MCFL only 5 remain. In contrast to the CFL case where several Goldstone bosons are Q-charged, in the MCFL all are Q-neutral. Therefore, the scattering rate of the low-energy bosons should be different in the magnetic background, and this will be reflected in turn in the transport properties of the star. By investigating transport properties as thermal conductivity and viscosity in the MCFL phase (or in the extension of the MCFL phase when one takes into account the strange quark mass and neutrality effects) one could look for new observational effects that will allow us to distinguish between nuclear-core stars and quark-core stars.

\underline{Thermal Conductivity}. Given that neutrino emission rates and heat capacity generally rise with density, neutron star cooling is likely preferentially sensitive to the properties of matter in the core. Investigations \cite{Shovkovy-Ellis-1, Shovkovy-Ellis-2} on the impact of the thermal conductivity of dense quark matter on the star cooling process indicates that any CFL quark matter within the star will cool by conduction, not by neutrino emission. In this direction, to investigate how the magnetic field can affect the medium thermal conductivity is of interest.

\underline{Neutrino emission and detection}. Neutrino emission is the dominant heat loss mechanism of the stars in their first million years. In \cite{Jaikumar-1, Jaikumar-2} the neutrino emission from Nambu-Goldstone modes of the CFL phase has been investigated. These studies showed that the scattering of massless Goldstone modes, associated with the breaking of the baryon $U(1)_B$ symmetry, is not exponentially suppressed, and so, these modes dominate neutrino emission at late times. On the other hand, the time-of-arrival distribution of supernova neutrinos could be connected to possible phase transitions to and in quark matter \cite{Alford-Jotwani, Carter}, but a detailed analysis of this suggestion requires a better understanding of both supernova itself and of the properties of quark matter at MeV temperatures.

If the CS phase at intermediate densities results to be a variety of gapless phase, the gapless modes could play a significant role in the transport properties of the star. A recent study \cite{Alford-Jotwani} has shown that even a relatively small region of gCFL matter in a star would dominate the heat capacity and the heat loss by neutrino emission. However, we already know that the gCFL is not stable, so it is unlikely that this phase will occur within the star.

However, none of these studies have taken into account the presence of the in-medium magnetic field that penetrates the star's superconducting core. Nevertheless, a total understanding of the transport mechanism of a compact star with a quark core will not be complete without considering the modification of the color-superconducting gap by the strong in-medium magnetic field, as well as by the modification of the remaining Goldstone modes.

This effect can be relevant for the low energy physics of a color
superconducting star's core and hence for its transport
properties. In particular, the cooling of a compact star is
determined by the particles with the lowest energy; so a star with
a core of quark matter and sufficiently large magnetic field can
have a distinctive cooling process. This study is a pending task that is worth to be undertaken.

\subsection{Boosting Stellar Magnetic Fields via an Internal Mechanism}
\label{subsec:2}

The standard model \cite{Magnetars-4,Magnetars-5}  to explain the origin of the strong magnetic fields observed in the surface of magnetars is based on a magnetohydrodynamic dynamo mechanism that amplifies a seed magnetic field due to a rapidly rotating protoneutron star. This model requires a spin period $<3 ms$.
Nevertheless, this mechanism cannot explain all the observed features of the supernova remnants surrounding these objects
\cite{magnetar-criticism-1, magnetar-criticism-2}.

As has been found recently, in color superconductors magnetic fields can be reinforced \cite{Vortex} and even generated  \cite{Vortex-Cond}. It is natural to expect that if a color superconducting state exists in the core of neutron stars, it may have implications for the magnetic properties of such compact objects. At the moderately high densities that exist in the cores of neutron stars the most probable color superconducting state is not the CFL,  but either an inhomogeneous phase or perhaps a strongly coupled 2SC phase. In the 2SC phase, the Meissner mass of the charged gluons decreases with decreasing density to values which are close to zero. For such small charged gluon masses, a magnetic field does not need to be too large to induce a vortex state. Fields with values $\widetilde{H}>m_M^2$ will trigger the spontaneous generation of vortices of charged gluons, which in turn will enhance the existing magnetic field. Hence,  CS could contribute to boosting the large magnetic fields observed in some stellar objects as magnetars,  without having to rely only on the quick spinning assumed in the standard model of magnetars that, as known, are associated with some of the conflict of this model with the observations. These induced gluon vortices could produce a magnetic field of the order of the Meissner mass scale, which implies a magnitude in the range $\sim 10^{16}-10^{17} G$. Hence, the possibility of generating a
magnetic field of such a large magnitude in the core of a compact
star without relying on a magnetohydrodynamics effect, can be an
interesting alternative to address the main criticism
\cite{magnetar-criticism-1, magnetar-criticism-2} to the observational conundrum of the
standard magnetar's paradigm \cite{Magnetars-1, Magnetars-2, Magnetars-3, Magnetars-4, Magnetars-5}. On the other hand, to
have a mechanism that associates the existence of high magnetic
fields with CS at moderate densities can serve to single out the
magnetars as the most probable astronomical objects for the
realization of this high-dense state of matter.

\subsection{Stability of Magnetized Quark Stars}
\label{subsec:3}

It is now our goal to analyze the conditions for
MCFL matter to become absolutely stable. This is done by comparing the
energy density at zero pressure condition with that of the iron nucleus ($\sim 930$ MeV). Depending on whether the energy density of the MCFL phase is higher or smaller than this value, the content of a magnetized strange quark could be or not made of MCFL matter.

The stability criterion for MCFL matter can be derived in a simple way. Following Farhi and Jaffe's \cite{Farhi84} approach, we can
determine the maximum value of the bag constant that satisfies
the stability condition at zero pressure for each magnetic field value. We call reader's attention that in all these derivations we work within a self-consistent approach, in which the solutions of the gap equations are substituted in the pressures and energies of each phase before imposing the conditions of equilibrium and stability.

After imposing the zero pressure condition in the EoS for the MCFL phase,
both the parallel
and perpendicular pressures in the MCFL EoS need to vanish
simultaneously. Therefore, the two equilibrium conditions become

\begin{eqnarray}
p^\|_{MCFL}&=&-\Omega_{MCFL}-{\cal{B}}-\frac{\widetilde{B}^2}{2}=0, \label{MCFL-Equilibrium-1}
\\
 p^\bot_{MCFL}&=&\widetilde{B} \frac{\partial\Omega_{MCFL}}{\partial \widetilde{B}}+\widetilde{B}
 \frac{\partial {\cal{B}}}{\partial \widetilde{B}}+\widetilde{B}^2=0  \label{MCFL-Equilibrium-2}
\end{eqnarray}

Where we are assuming that the bag constant depends on the magnetic
field. It is not unnatural to expect that the applied magnetic
field could modify the QCD vacuum, hence producing a field-dependent bag constant. One can readily
verify that Eqs.
(\ref{MCFL-Equilibrium-1})-(\ref{MCFL-Equilibrium-2}) are
equivalent to require $p^\|_{MCFL}=0$ and $\partial
p^\|_{MCFL}/\partial \widetilde{B}=0$ at the equilibrium point.

Equation (\ref{MCFL-Equilibrium-2}) can be rewritten as
\begin{equation}  \label{Equilibrium-3}
\widetilde{B}=M-\frac{\partial {\cal{B}}}{\partial \widetilde{B}}
\end{equation}
where $M=-\partial \Omega_{MCFL}/\partial \widetilde{B}$ is the
system magnetization. If we were to consider that the vacuum energy
${\cal{B}}$ does not depend on the magnetic field, we would need
\begin{equation}  \label{Magnetization}
M=\widetilde{B},
\end{equation}
to ensure the equilibrium of the self-bound matter, a condition difficult to satisfy since
it would imply that the medium response to the applied magnetic
field (i.e. the medium magnetization $M$) is of the order of the
applied field that produces it. Only if the MCFL matter were a ferromagnet this would be viable, but as known, the MCFL matter is on the contrary, an insulator. The other possibility for the
equilibrium conditions (\ref{MCFL-Equilibrium-1}) and
(\ref{MCFL-Equilibrium-2}) to hold simultaneously is to have a
field-dependent bag constant capable to yield nonzero vacuum magnetization
$M_0=-\frac{\partial {\cal{B}}}{\partial \widetilde{B}}\simeq \widetilde{B}$.

\begin{figure}
\begin{center}
\includegraphics[width=0.44\textwidth]{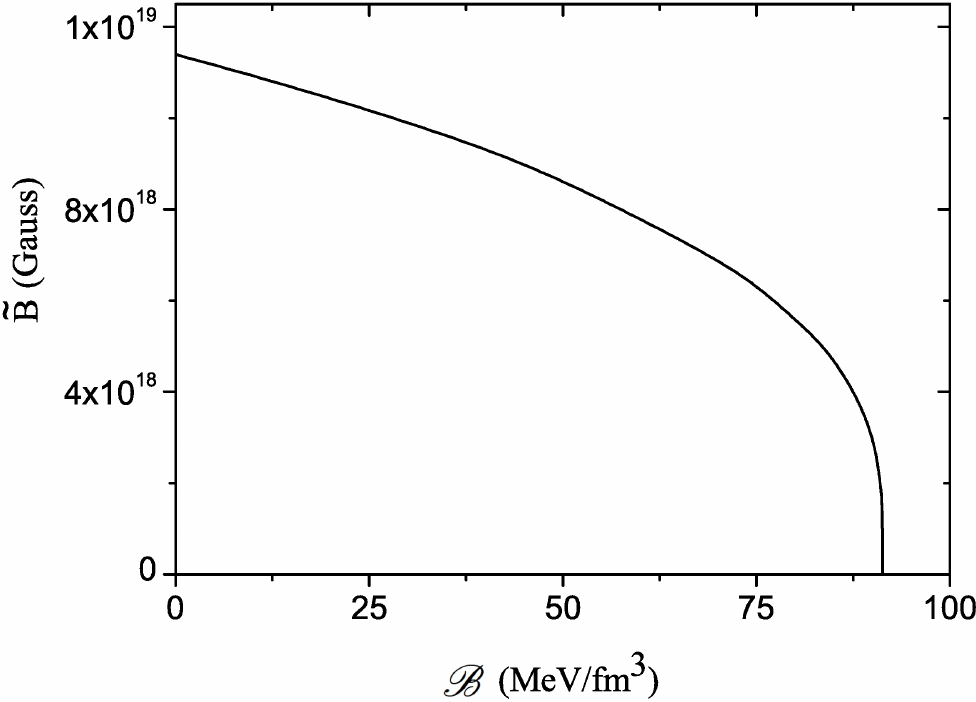}
\caption{\footnotesize Stability window for MCFL matter in the
plane $\tilde{B}$ vs. ${\cal{B}}$. The curve shown corresponds to the
borderline value $\epsilon/A=930$ MeV.} \label{window}
\end{center}
\end{figure}

The following comment is in order. The fact that the bag constant
needs to be field-dependent for self-bound stars in a strong magnetic field is a direct consequence of the lack of
a compensating effect for the internal pressure produced by the
magnetic field other than that applied by the vacuum (an exception could be of course if the paired quark matter would exhibit ferromagnetism). For gravitationally bound stars,
on the other hand, the situation is different, since the own
gravitational field can supply the pressure to compensate the one
due to the field. For such systems, keeping ${\cal{B}}$ constant in
the EoS is in principle possible. Under this assumption we
considered a fixed ${\cal{B}}$-value in Fig. \ref{EOS}.

Taking into account that the matter energy density $\epsilon^{'}_{MCFL}$ (i.e. the energy density that does not include the pure Maxwell contribution) divided by the baryon number is given by
\begin{equation}\label{Matter-energy}
\frac{\epsilon^{'}_{MCFL}}{n_A}=\frac{\Omega_{MCFL}+{\cal{B}}}{n_A}-\frac{\mu}{n_A}\frac{\partial\Omega_{MCFL}}{\partial \mu},
\end{equation}
We can write it under the zero parallel pressure condition ($\Omega_{MCFL}+{\cal{B}}=-\widetilde{B}/2$) as
\begin{equation}\label{Matter-energy-2}
\frac{\epsilon^{'}_{MCFL}}{n_A}|_{\mu_B}=2\frac{\widetilde{B}^2}{2n_A}+\frac{\mu_{B}}{n_A}N,
\end{equation}
and the absolute stability condition becomes
\begin{equation}\label{Matter-energy-3}
\frac{\epsilon^{'}_{MCFL}}{n_A}|_{\mu_B}=2\frac{\widetilde{B}^2}{2n_A}+\frac{\mu_{B}}{n_A}N \leq \epsilon_{0}(Fe^{56}),
\end{equation}

Then, finding $\mu_B$ as a function of $\widetilde{B}$ from (\ref{Matter-energy-3}), and substituting it back in (\ref{MCFL-Equilibrium-1}), we can numerically solve
\begin{equation}\label{bagH}
{\cal{B}}(\widetilde{B})=-\Omega_{MCFL}(\mu_B,\widetilde{B})-\widetilde{B}^2/2,
\end{equation}
to determine the stability window in the plane
$\widetilde{B}$ versus ${\cal{B}}$ for the MCFL matter to be absolute stable
(Fig. \ref{window}). The
inner region, which corresponds to smaller bag constants for each given
$\widetilde{B}$, is the absolutely stable region.

Note that, contrary to Farhi and Jaffe \cite{Farhi84}, we did not impose a {\it minimum} value for the bag
constant because we have no clear indication from experiments of the
possible behavior of this parameter when a magnetic field is
applied to a system.

In summary, our results indicate that a condition for the MCFL matter to be absolutely stable is a field-dependent bag constant that can give rise to a large vacuum magnetization at moderately strong fields (see Fig. \ref{window}). Under these circumstances, increasing the magnetic field tends to destabilize the self-bound MCFL matter. This result differs from that found in \cite{Aurora} where it was used a CFL model at $eB \neq 0$ with only one
gap that was fixed by hand.

\begin{acknowledgement}
This work has been supported in part by DOE Nuclear Theory grant DE-SC0002179.
\end{acknowledgement}

\end{document}